\documentclass{emulateapj}
\usepackage{graphicx,apjfonts}
\usepackage{natbib}
\usepackage{lscape}

\slugcomment{Accepted by ApJ on Oct.25th 2008}

\shorttitle{}
\shortauthors{}
\begin{document}

\title{Mergers of luminous early-type galaxies in the local universe and
gravitational wave background}

\author{Z. L. Wen\altaffilmark{1,3}, 
        F. S. Liu\altaffilmark{2,1}, 
        and 
        J. L. Han\altaffilmark{1}}

\altaffiltext{1}{National Astronomical Observatories, Chinese Academy of Sciences, 
                 20A Datun Road, Chaoyang District, Beijing 100012, P.R.China; 
                 zhonglue@bao.ac.cn, hjl@bao.ac.cn.} 
\altaffiltext{2}{College of Physics Science and Technology, 
                 Shenyang Normal University, Shenyang 110034, P.R.China; 
                 lfs@bao.ac.cn.}
\altaffiltext{3}{Graduate University of the Chinese Academy of Sciences, 
                 Beijing, 100049, P.R.China}

%%%%%%%%%%%%%%%%%%%%%%%%%%%%%%%%%%%%%%%%%%%%%%%%%%%%%%%%%%%%%%%%%%%%%%%%%%%%%

\begin{abstract}
Supermassive black hole (SMBH) coalescence in galaxy mergers is
believed to be one of the primary sources of very low frequency
gravitational waves (GWs). Significant contribution of the GWs comes from
mergers of massive galaxies with redshifts $z<2$. Very few previous
studies gave the merger rate of massive galaxies.
We selected a large sample (1209) of close pairs of galaxies with projected
separations $7<r_p<$50 kpc from 87,889 luminous early-type galaxies
($M_r<-21.5$) from the Sloan Digital Sky Survey Data Release
6. These pairs constitute a complete volume-limited sample in the
local universe ($z<0.12$). Using our newly developed technique, 249
mergers have been identified by searching for interaction
features. From them, we found that the merger fraction of luminous
early-type galaxies is 0.8\%, and the merger rate in the local
universe is
$R_{\rm g} \sim (1.0\pm0.4)\times 10^{-5}~{\rm ~Mpc^{-3}~Gyr^{-1}}$
with an uncertainty mainly depending on the merging timescale.
We estimated the masses of SMBHs in the centers of merging galaxies based
on their luminosities. We found that the chirp mass distribution of
the SMBH binaries follows a power law with an index of $-3.0\pm0.5$ in the
range $5\times 10^8$--$5\times 10^{9}$~$M_{\odot}$.
Using the SMBH population in the mergers and assuming that the SMBHs
can be efficiently driven into the GW regime, we investigated the
stochastic GW background in the frequency range $10^{-9}$--$10^{-7}$ Hz. 
We obtained the spectrum of the GW background of
$h_c(f)\sim 10^{-15}(f/{\rm yr}^{-1})^{-2/3}$, which is one magnitude
higher than that obtained by Jaffe \& Backer in 2003, but consistent
with those calculated from galaxy-formation models.

\end{abstract}


%%%%%%%%%%%%%%%%%%%%%%%%%%%%%%%%%%%%%%%%%%%%%%%%%%%%%%%%%%%%%%%%%%%%%%%
\section{Introduction}

Gravitational waves (GWs) are a new window to observe violent
astrophysical dynamic processes. Efforts in the detection of GWs are
currently being made at several frequency bands.
For example, the Laser Interferometer Gravitational Wave Observatory
(LIGO) aims to detect GWs in the frequency range 1--$10^4$~Hz
\citep{aad+92} emitted from the coalescence of binary compact objects
or from the cosmic GW background.
In the frequency range $10^{-6}$--0.1~Hz, the Laser
Interferometer Space Antenna (LISA) is expected to detect the GWs
emitted from the coalescence of massive black hole (BH) binaries in the
mass range $10^3$--$10^7$ $M_{\odot}$ and from the unresolved
white dwarf binaries in the Milky Way and nearby galaxies
\citep{hae94,nyp01}.
Precision timing of millisecond pulsars appears to be a unique
technique \citep{saz78,det79,jhl+05,jhv+06} to measure GWs emitted
from the coalescence of supermassive black holes (SMBHs) with masses 
$10^7$--$10^{10}~M_{\odot}$ in galactic nuclei, cosmic superstrings, and
relic GW background from the big bang \citep{mag00,phi01,jb03} in the
frequency range $10^{-9}$--$10^{-7}$~Hz.
  
The SMBH mergers at redshifts $z<2$ dominate the GW background and
resolvable signal at a frequency band $10^{-9}$--$10^{-7}$~Hz
\citep{wl03,shmv04,svc08,svv08}. Only galaxy mergers provide a chance 
for the SMBH mergers. At present, the galaxy merger history is
not very clear. The evolution of the merger rate is often described by a
power law, $R_{\rm g}(z)\propto (1+z)^m$, where, $m$ is the evolution
index. However, the merger rate of massive galaxies has not been
determined precisely from observations. In the rest of the introduction,
we will review the current knowledge on galaxy pair fraction and
merger rate, and then discuss current understanding on GWs from
SMBH coalescence.

In this paper, we assume a $\Lambda$CDM cosmology taking
$H_0=$100 $h$ ${\rm km~s}^{-1}$ ${\rm Mpc}^{-1}$, with $h=0.72$,
$\Omega_m=0.3$ and $\Omega_{\Lambda}=0.7$.

\subsection{Galaxy pair fraction and merger rate: current knowledge}

Three methods have previously been applied for measurement of the 
merger rate and its evolution.
The first and most straightforward is to count the incidence of
strong disturbed galaxies with double nucleus, or 
tidal tails \citep[e.g.][]{lal+00,
  cbd+03,ccw03,lrc+04,van05,klc+07,ldf+08}.
The second is to take statistics of close pairs from a 
spectroscopic redshift survey assuming that the close pairs will
result in galaxy mergers over a relatively short timescale
\citep{zk89,bkw+94,cpi94,ye95,wfr95,nir+97,ppy+97,wk98,pcm+00,
  ppc+02,bfe+04,lkw+04,dld+05,dcl+07,bnm+06,kss+07,lpk+08,pa08}. Physical
pairs with separations $r_p<$20 $h^{-1}$~kpc and line-of-sight
velocity differences $\Delta v<500$ ${\rm km~s}^{-1}$ are expected to
merge within 0.5~Gyr \citep{pcm+00, con06}. Key problems of such
studies were the very limited number of galaxies with spectroscopic
redshifts, the incompleteness of the sample, and the
contamination of unphysical pairs \citep{pcm+00,dcl+07}.
Another useful but less direct method is the statistics of close pairs
from the two-point correlation function of galaxies at small scales
\citep[e.g.][]{mhc+06,bps+06}. The projected correlation function can
be obtained from a large sample of galaxies, and can be
``deprojected'' to get the real-space correlation function.

Morphological signatures of interactions can be found by visual
inspection, which is very time consuming and somewhat subjective.  Some
excellent studies have been published recently.
\citet{lal+00} studied merger fraction using 285 galaxies from the
Canada--France Redshift Survey and Autofib-Low Dispersion Spectrograph
Survey. They found 49 pairs by visual identification, 37 of which
have a Lee ratio $L_R>1.5$. Combining the result given by
\citet{ppy+97}, \citet{lal+00} determined the merger fraction
evolution as $2.1\% \times (1+z)^{3.4\pm0.6}$.
\citet{cbd+03} studied the fraction of galaxies undergoing major
mergers as a function of redshift using model-independent
morphological signatures (concentration, asymmetry, clumpiness) in the
WFPC2 and NICMOS Hubble Deep Field North. Their samples have 51, 142,
93, and 183 galaxies ($M_B<-18$) in the four redshift ranges with $\langle
z \rangle =$ 0.58, 1.10, 1.73, and 1.41, respectively. They found that
the corresponding merger fractions are 4\%, 14\%, 14\%, and 9\%.
Recently, \citet{ldf+08} found 312 merger candidates with
morphological disturbances based on a volume-limited sample of 3009
galaxies ($M_B<-18.94$) from the All-Wavelength Extended
Groth Strip International Survey. A constant merger fraction of
(10$\pm$2)\% was found in the redshift range of $z=0.2$--1.2.

There has been much effort applied to the statistics of close pairs, as
summarized in Table 2 of \citet{kss+07}.
\citet{ccp+00} selected kinematic pairs of galaxies ($M_B-5\log
h<-19.8$) with $5<r_p<100$ $h^{-1}$~kpc and $\Delta v<1000$ ${\rm
  km~s}^{-1}$ in the redshift range of $z=0.1$--1.1. They found 18
pairs from 300 galaxies of the Caltech Faint Galaxy Redshift Survey and 91
pairs from 3000 galaxies of the Canadian Network for Observational
Cosmology.
\citet{pcm+00} defined $N_c$, ``the number of dynamically close {\it
  companions per galaxy}'', to study the pair fraction. From 5426
galaxies ($-21<M_B<-18$) in the Second Southern Sky
Redshift Surveys, they found 80 {\it companions} satisfying the
conditions $5<r_p<20$ $h^{-1}$~kpc and $\Delta v<500$ ${\rm
  km~s}^{-1}$, and derived $N_c=(2.26\pm0.52)\%$ at an average
redshift $\langle z \rangle=0.015$.
Using the same selection criteria, \citet{ppc+02} identified 88
galaxies in close pairs from 4184 field galaxies at redshifts
$0.12\leq z \leq 0.55$ from the Canadian Network for Observational
Cosmology. They obtained $N_c = (3.21\pm0.77)\%$ at redshift $\langle
z \rangle=0.3$. Combing the early result in \citet{pcm+00}, they also
determined the merger rate evolution as $(1+z)^{2.3\pm0.7}$.
\citet{lkw+04} found 79 paired galaxies ($10<r_p<50$ $h^{-1}$~kpc,
$\Delta v<500$ ${\rm km~s}^{-1}$) out of 2547 galaxies from the
initial data of the DEEP2 Redshift Survey, and found that the pair
fraction $N_c$ is $\sim8\%$ at redshift $z\sim0.6$ and increases to
$\sim10\%$ at redshift $z\sim1.1$. Fitting a power-law model to the
pair fractions determined from higher redshift together with those
determined at lower redshift \citep{pcm+00,ppc+02}, \citet{lkw+04} obtained 
the evolution index $m=0.51\pm0.28$.
From the Millennium Galaxy Catalogue, \citet{dld+05} found 137
dynamically close companions ($5<r_p<20$ $h^{-1}$~kpc, $\Delta v<500$
${\rm km~s}^{-1}$) in a bright sample ($-22 < M_B-5\log h < -19$) and
176 companions in a faint sample ($-22 < M_B-5\log h < -18$). They
found the pair fractions $N_c=(1.74\pm0.15)\%$ at redshift $\langle z
\rangle = 0.123$ and $(3.57\pm0.27)\%$ at redshift $\langle z \rangle
=0.116$, respectively, after a correction for ($\sim30\%$) missing
pairs.

Above studies have used relatively small samples. Recently, 1749 close
pairs ($5<r_p<20$ kpc) have been found by \citet{kss+07} from a
complete sample of 59,221 galaxies ($M_V<-19.8$) in the redshift range
of $z=0.1$--1.2 from the Cosmic Evolution Survey field. This is the
largest data set of close pairs.  Supplemented by the local pair
fraction from the Sloan Digital Sky Survey (SDSS), they found the pair
fraction evolution to be $(1+z)^{3.1\pm0.1}$.
Note that the majority of these statistics were studied for general
mergers, using close pairs selected with small radial velocity
differences and small projected separations.

Recently, several studies have been carried out on mergers between
gas-poor early-type massive galaxies, also called ``dry
merger'' \citep{bnm+06,lpk+08}.
\citet{mhc+06} studied the merger rate of Luminous Red Galaxies in the
SDSS using the two-point correlation function. They found that the
correlation function closely follows $\xi(r)\sim r^{-2}$ over four orders
of magnitude, from 0.01 to 100~$h^{-1}$~Mpc. Taking a merger length
scale of $r_f=10$~kpc and a typical velocity dispersion, $\sigma_v\sim
200$ ${\rm km~s}^{-1}$, the dynamical time is about $t_{\rm dyn}=
$200~Myr. The merger rate per galaxy, $\Gamma$, for galaxies with a
comoving number density, $n=10^{-4}~{\rm Mpc^{-3}}$, is
$$\Gamma \approx \frac{4\pi n\; r_f^2\xi(r_f)}{t_{\rm
dyn}}=\frac{1}{160~{\rm Gyr}}. $$
We can further find a comoving volume merger rate of galaxies,
\begin{equation}
R_{\rm g}\equiv n\Gamma=6\times 10^{-7}~{\rm Mpc}^{-3}~{\rm Gyr}^{-1}.
\end{equation}
Note that this is one magnitude smaller than that of
\citet{mhc+06}.\footnote{Equation (12) in \citet{mhc+06} has a wrong form
  of $\Gamma/n$.}
\citet{bnm+06} found six dry mergers (12 galaxies) from 468 early-type
galaxies ($M_V<-20.5$) in the redshift range of $z=0.1$--0.7 from
the Galaxy Evolution from Morphology and spectral energy distributions (SEDs) survey. Their
simulations show that {\it the distinct interaction features in the
  mergers of early-type galaxies only appear in the last pass or
  coalescence}.
Following this, \citet{mgh+08} identified 38 merging pairs of massive
galaxies (stellar masses $M_{\rm star} >5\times 10^{10}$ $M_{\odot}$
and $r_p\le30$~kpc) from 845 galaxy groups/clusters at redshifts
$z<0.12$ in the SDSS Data Release 2 (DR2). They found that the merger rate
of massive galaxies in the galaxy groups is several times higher than
that of the SDSS Luminous Red Galaxies.
\citet{lpk+08} studied the evolution of pair fraction and merger rate
for different types of pairs of galaxies ($-21 < M_B < -19$) with
$10<r_p< 30$, 50, 100 $h^{-1}$~kpc and $\Delta v<500$ ${\rm
  km~s}^{-1}$. Their sample includes 218 blue--blue pairs, 122 red--red
pairs and 166 red--blue pairs in the redshift range of $z= 0.1$--1.2
from Team Keck Redshift Survey and other surveys mentioned above.
They found that the merger rate evolutions are different for
different types of pairs.  The blue--blue pairs have a merger rate
evolution index $m=1.27\pm0.35$, whereas the red--red pairs and
red--blue pairs have negative indices, as $-0.92\pm0.59$ and
$-1.52\pm0.42$, respectively.

In summary, most previous studies work on the pair fraction and merger
rate of a general merger. The pair fraction varies in a range of 1\%--10\% 
in the redshift range of $z= 0.2$--1.2 with an evolution index
of $m=0-3$. Only a few authors have tried to determine merger rates
for different types of galaxies. The mergers of early-type galaxies have
been explored only recently \citep{bnm+06, mhc+06,lpk+08,ldf+08}. The
large uncertainty of merger rate remains due to either the small
sample or the contamination of unphysical pairs. The merger rate
should be better determined using the fraction of merging galaxies,
rather than the fraction of galaxies in the projected close pairs.

Using the SDSS data, we select a large complete volume-limited 
pair sample of luminous early-type galaxies and try to determine the merger
rate in the local universe ($z<0.12$). The BHs in the luminous
early-type galaxies are much more massive than those in late-type
galaxies, and their mergers will play an important role in the
formation of massive galaxies and SMBH binaries.

\subsection{SMBH mergers and GWs}

SMBHs exist in the nucleus of nearby and distant galaxies.  When
two galaxies merge, the SMBHs in the centers of galaxies sink toward the
center of a newly formed galaxy through dynamic friction
\citep{bbr80,yu02}, and then a SMBH binary can be formed. The SMBH
binary continues to harden (i.e., lose energy, shrink, and move faster)
and becomes a bound system through, e.g., stellar dynamics
\citep{qui96,mm01,shm06} or gas dynamics
\citep{gr00,an02,elc+04,dch+07}. Consequently, the SMBH binary loses
energy and angular momentum by GW radiation, and eventually coalesces 
to produce a luminous GW event.

The GWs emitted from SMBH binaries can be detected as individual
sources and a stochastic background from many events together. Recent
studies \citep{shv+88,simt03,vkl+08} have suggested that SMBH binaries
may exist in the centers of galaxies with orbit periods of a few
years. They are possible individual GW-emitting sources to be revealed
by pulsar timing observations.  The GW amplitude from a SMBH binary is
\citep{tho87}
\begin{equation}
h_s=4\sqrt{\frac{2}{5}}\frac{(GM_c)^{5/3}}{c^4D(z)}[\pi f(1+z)]^{2/3},
\label{hs}
\end{equation}
where $M_c$ is the chirp mass of the SMBH binary,
\begin{equation}
M_c^{5/3}=\frac{M_1M_2}{(M_1+M_2)^{1/3}};
\label{chirpmass}
\end{equation}
here $M_1$ and $M_2$ are SMBH masses; $f$ is the observed GW
frequency; $c$ is the speed of light; $D(z)$ is the comoving distance
to the system located at $z$,
\begin{equation}
D(z)=\frac{c}{H_0}\int_0^z \frac{dz}{E(z)};
\end{equation}
here $E(z)=\sqrt{\Omega_{\Lambda}+\Omega_m(1+z)^3}$. The frequency
change of the GW per unit time in the observer's frame is given by
\citet{pm63}:
\begin{equation}
\frac{df}{dt}=\frac{96}{5\pi}\Big(\frac{GM_c}{c^3}\Big)^{5/3}(\pi
f)^{11/3}(1+z)^{5/3}.
\label{timeeq}
\end{equation}

A large number of unresolved coalescing SMBH binaries together can produce a
stochastic GW background. The spectrum can be formulated as
\citep{phi01,jb03,eins04}
\begin{equation}
h_c^2(f)=\int dz\;dM_c\;  h_s^2  N(f,z,M_c) \;
\; f \; \theta(f_{\rm max}-f),
\label{hcf}
\end{equation}
where $N(f,z,M_c)\;dz\;dM_c$ is the number of SMBH binaries per unit
frequency in mass interval $dM_c$ and redshift interval $dz$.  The
$\theta(x)$ in Equation~\ref{hcf} is the step function \citep{eins04} and
$f_{\rm max}=c^3/[6^{3/2}\pi G(M_1+M_2)]$ is the maximum frequency of the 
GW before the SMBHs plunge together \citep{hug02}. $N(f,z,M_c)$ is
related to the merger rate of SMBH per unit comoving volume, $R_{\rm
  BH}(z)$, by 
\begin{equation}
N(f,z,M_c)=\frac{4\pi c^3}{H_0^3}\frac{D(z)^2}{E(z)(1+z)}
\frac{dt}{df}\frac{\Phi(M_c,z)}{n_{\rm BH}(z)}R_{\rm BH}(z),
\label{numbin}
\end{equation} 
here, $\Phi(M_c,z)$ is the chirp mass distribution of SMBH binaries;
$n_{\rm BH}(z)=\int \Phi(M_c,z)\;dM_c$ is the number density of SMBH
binaries.

\citet{rr95} estimated the number of GW-emitting sources by examining
the probability of SMBH binary coalescence through the process by
interaction with field stars. They used the SMBH mass function from
the model of active galactic nuclei \citep[AGNs;][]{sb92} and a merger rate evolution as
$(1+z)^{3.5}$. Finally, they obtained a GW spectrum of $h_c(f)\sim
10^{-16}(f/\rm yr^{-1})^{-2/3}$ at frequency $f\sim 0.1-1 {\rm
  yr}^{-1}$.
\citet{phi01} further formulated the calculation of the GW background,
and obtained the spectrum in a power law $h_c(f)\sim f^{-2/3}$. He
found that the GW spectrum index is independent of cosmology. The
amplitude is in an order of $\sim 10^{-16}$ at the frequency of $1
{\rm yr}^{-1}$ which depends on the merger rate of SMBHs and the SMBH
population in the universe.

\citet{jb03} formulated all ingredients including the merger rate of
SMBHs, the distribution of SMBH masses, the strain of the GW
background for a single binary, and the GW radiation timescale
($\tau_{\rm GW}=f\;dt/df$). They obtained the spectrum\footnote{The
  authors misquoted the strain amplitude of $h_c(f)\sim 10^{-15}(f/\rm
  yr^{-1})^{-2/3}$ in their abstract.} of $h_c(f)\sim 10^{-16}(f/\rm
yr^{-1})^{-2/3}$, confirming the results of \citet{rr95} and
\citet{phi01}. They also found that the slope of the spectrum is
determined by the strain and timescale of individual merger event,
independent of the merger models and SMBH population. The large
uncertainty of the spectrum amplitudes comes from approximations in
the theoretical formulation, lack of knowledge of the merger rate and
SMBH population, and unknown dynamical processes of SMBH binary to
reach a GW-dominated regime.

To minimize the uncertainties, some efforts have been made to
calculate the GW background utilizing the galaxy merger rate and the
SMBH population based on the standard hierarchical structure formation
scenarios \citep{wl03,eins04,shmv04,svc08}. These studies show that
the amplitude of the GW background is a few times higher than those
given by \citet{phi01} and \citet{jb03}. The characteristic strain
spectrum is dominated by SMBHs with masses larger than $10^{9}$
$M_{\odot}$ at low redshifts ($z<2$), and the spectrum becomes steeper
than $-2/3$ when the frequency is larger than about $10^{-7}$~Hz.

Clearly, observational determination of the merger rate of SMBHs and
the SMBH mass function in the binaries is crucial for a more precise
calculation of the GW background. This paper is organized as
follows. In Section 2, we describe the identification of merging pairs and
determine the merger rate of luminous early-type galaxies in the local
universe. In Section 3, we derive the chirp mass distribution of the SMBH
binaries in the mergers and calculate the spectrum amplitude of the GW
background. We give conclusions in Section 4.

%%%%%%%%%%%%%%%%%%%%%%%%%%%%%%%%%%%%%%%%%%%%%%%%%%%%%%%%%%%%%%%%%%%%%%%%%%%%
%
\section{Pair fraction and merger rate of luminous early-type galaxies in the SDSS}
%
%%%%%%%%%%%%%%%%%%%%%%%%%%%%%%%%%%%%%%%%%%%%%%%%%%%%%%%%%%%%%%%%%%%%%%%%%%%%%%

To determine the pair fraction and merger rate of luminous early-type
galaxies in the local universe, we selected a large and complete
sample of close pairs of galaxies ($M_r<-21.5$) at redshifts $z<0.12$
directly from the SDSS DR6. The SDSS provides
photometric data in five broad bands ($u$, $g$, $r$, $i$, and $z$) for
more than 8400 deg$^2$ to the limit of $r=22$ mag, deeper than
any previous wide sky surveys. The follow-up spectroscopy observations
have measured spectra of more than 790,000 galaxies in about 7425
deg$^2$ \citep{dr6}. The main galaxy sample reaches an
extinction-corrected Petrosian magnitude of $r=17.77$ \citep{swl+02}
and a completeness of $\sim$90\% \citep{bll+03}. We selected this
sample of projected close pairs incorporating the spectroscopic with
photometric catalogs, and then identified the mergers by searching for
the interaction features.

\subsection{Pairs of luminous early-type galaxies: sample}

In previous studies, close pairs were usually selected using the
criteria $\Delta v<500$ ${\rm km~s}^{-1}$ and $r_p<20$ $h^{-1}$~kpc
or $r_p<30$ $h^{-1}$~kpc (see Section 1.1). However, direct searches of
pairs from the SDSS spectroscopic catalog would miss $\sim$70\% of
close pairs due to the fiber collision problem \citep{mgh+08}. Targets of
the spectroscopic survey are assigned to fibers with a radius of 1.5
arcsec, and two fibers cannot be placed more closely than 55
arcsec. The incompleteness is severe on very small angular
scales. However, the estimated photometric redshifts \citep{olc+08}
can be used as the complement of the spectroscopic redshifts of
galaxies.

We obtained a complete pair sample of luminous early-type galaxies
from the SDSS with the following steps.  We found the redshifts of
bright galaxies ($13.5<r<17.5$) and constructed a sample of close
pairs with projected separations $7<r_p<50$~kpc at redshifts
$z<0.12$. The photometric redshifts are taken for the galaxies without
spectroscopic redshifts. If the spectroscopic redshifts are available
for both galaxies in a pair (i.e., the spec--spec pair), {\it the
  redshift for the pair} is taken as the average of the two
redshifts. If the spectroscopic redshift is available for only one of
the paired galaxies (i.e., the spec--phot pair), it is taken as the
pair redshift. If no spectroscopic redshifts are available for both
galaxies (i.e., the phot--phot pair), the pair redshift has to be taken
as the average of the two photometric redshifts.
The lower limit of 7~kpc in separation corresponds to $\sim3''$ at
redshift $z=0.12$. The photometry becomes unreliable for paired
galaxies with smaller angular separations \citep{mhc+06}. The upper
limit of 50~kpc is chosen so that all merging pairs, even the one with 
a large separation, can be included in our sample. We restricted the
redshift range of $z<0.12$ to get enough samples for our statistics and
to have the galaxies well resolved for the further image analysis.
We applied the $K$-correction for all galaxies to the rest frame
\citep{br07}. The galaxies in the pairs must have $M_r<-21.5$ and
satisfy the color constraints of $(u-r)>2.2$ and $(g-r)>0.7$
\citep{sik+01}. However, the sample is still contaminated by some red
late-type galaxies, e.g., edge-on spiral galaxies with dust lanes or
galaxies with red bulges, which were about $10\%$ and have been
excluded by visual inspection on color images of all targets from the
{\it DR6 Catalog Archive
  Server}\footnote{http://cas.sdss.org/astro/en/}. Following the above
selection criteria, we obtained 1209 pairs from 87,889 galaxies with
$M_r<-21.5$ at redshifts $z<0.12$. There are 230 spec--spec pairs, 543
spec--phot pairs and 436 phot--phot pairs.  These pairs are listed in
Table 1 (a full list is available in the online version).

\begin{figure}[ht!]
\epsscale{1.1}
\plotone{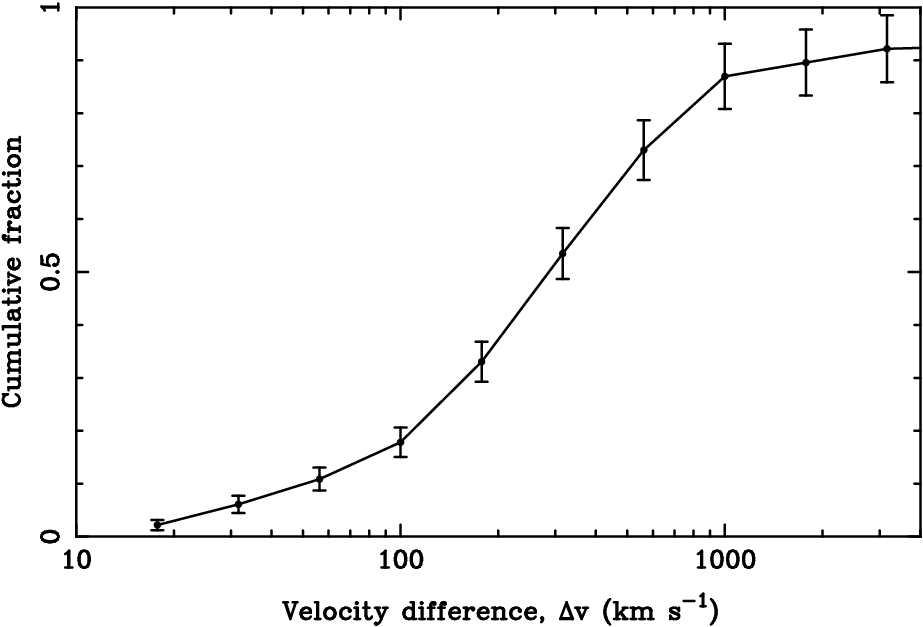}\\[2mm]
\plotone{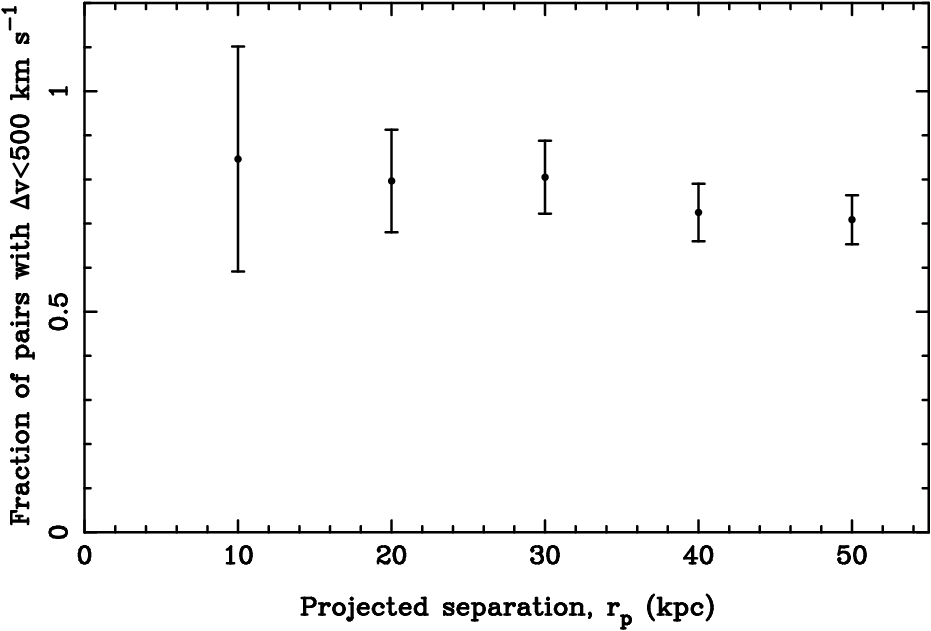}
\caption{Top: distribution of line-of-sight velocity differences
  for the spec--spec pairs. Bottom: fractions of pairs with
  velocity differences $\Delta v < 500$ ${\rm km~s}^{-1}$ for various 
  maximum projected separations.
\label{velhist}}
\end{figure}

Figure~\ref{velhist} shows the distribution of line-of-sight velocity
differences for the spec--spec sample. 71\% of the pairs have 
$\Delta v<500$ ${\rm km~s}^{-1}$. We also show in
Figure~\ref{velhist} the fraction of pairs with $\Delta v < 500$ ${\rm
  km~s}^{-1}$ as a function of projected separation. Among the close
pairs with $r_p<30$ kpc, 81\% have the $\Delta v < 500$ ${\rm
  km~s}^{-1}$.

\subsection{Identification of mergers}

Most of the previous efforts in the determination of the merger rate used the
projected close pairs. Very few groups \citep{bnm+06,mgh+08}
identified interaction features for pairs of luminous early-type
galaxies. The apparent close pairs may be widely separated in a 
three-dimensional space. It is necessary to check how many of the projected
close pairs are merging.

\begin{figure}[ht]
\epsscale{1.1}
\plotone{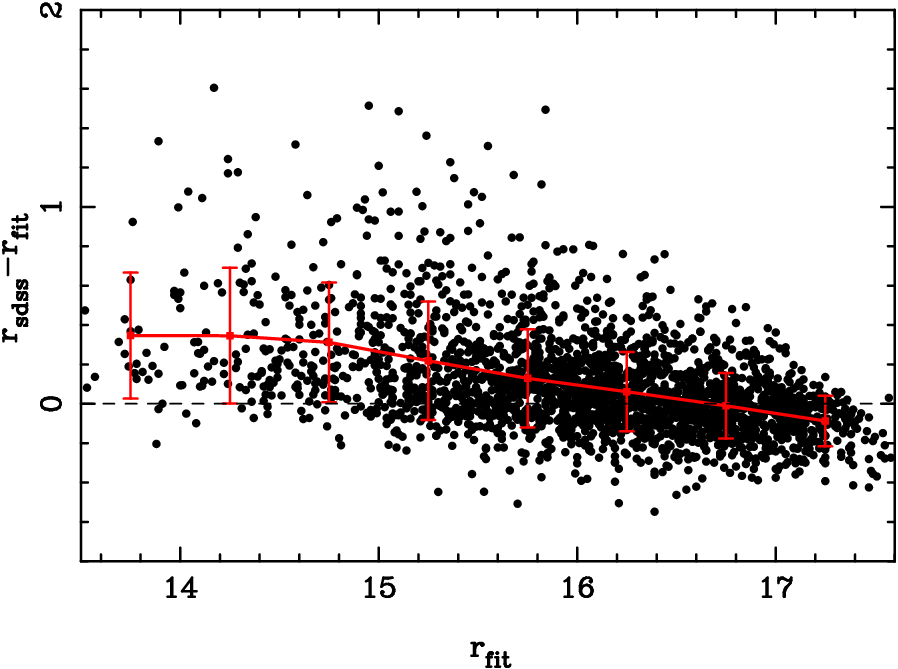}
\caption{Difference between the SDSS model magnitude and our fitted
  magnitude against the fitted magnitude for paired galaxies.
\label{magcom}}
\end{figure}

In merging galaxies, stars and gas may be torn apart from parent
galaxies under tidal force. The induced asymmetric features (such as
tails, bridges and plumes) can be used as evidence for galaxy
interactions.
Interaction involving at least one gas-rich late-type galaxy usually
accompanies strong star forming and creates distinct long tidal
tails. Such a merger is distinct and easy to identify. In
contrast, merger involving only gas-poor early-type galaxies usually
does not create distinct interaction features. However, the weak
features of early-type galaxy merging can be identified after a smooth
symmetric model for each paired galaxy is subtracted from the image
\citep{bnm+06,mgh+08}.

We extracted the SDSS $r$-band images for all selected pairs. The
corrected frames have been processed by the SDSS pipeline, including
bias-subtraction, flat-field, cosmic-ray removal, and correction for
pixel defect. We performed a precise sky background subtraction from
the corrected frames \citep[see details in][]{lxm+08}.
We applied the GALFIT package \citep{phir02} to construct a smooth
symmetric model for every early-type galaxy in the projected pair
list. The model fitting requires an observational image, a masked
image, and an image of the point-spread function (PSF) that is the
seeing of SDSS images characterized by the parameters of
double-Gaussian profiles \citep{slb+02}.
From the SDSS catalog, we also obtained a list of other fainter
galaxies and stars within $2R_{90}$ from the centers of two
galaxies. Here $R_{90}$ is the radius containing 90\% of the Petrosian
flux. The stars have directly been modeled with the PSF, but the
target early-type galaxies and other fainter galaxies have been
modeled by the S{\'e}rsic function \citep{ser68} convolved with the
PSF image. Objects {\it outside} $2R_{90}$ in the extracted image were
masked. The fitting is to minimize the $\chi^2$ between the
sky-subtracted image of unmasked pixels and the PSF-convolved model
for paired galaxies and other objects in the image.

The modeling procedure provides a fitted magnitude for each galaxy,
which is the integrated flux of the best-fitted S{\'e}rsic
function. The advantage of the fitted magnitude is that the fluxes
from both galaxies in the overlapped region can be
separated. Moreover, we have corrected the sky background for bright
objects and objects in crowded fields in the SDSS pipeline \citep[see
  more detailed discussion and solution in][]{lxm+08}, which would
result in a systematic underestimation for luminosities (and sizes)
of bright objects in the SDSS pipeline
\citep{mhs+05,lfr+07,lxm+08}. Brighter galaxies tend to be influenced
more severely \citep[see Figure 3 of][]{lxm+08}. This tendency can be
seen in Figure~\ref{magcom} from a rough comparison between the fitted
magnitudes and the SDSS model magnitudes. Therefore, the fitted
magnitudes are adopted in the following analysis.

\begin{figure}[t]
\epsscale{1.15}
\plotone{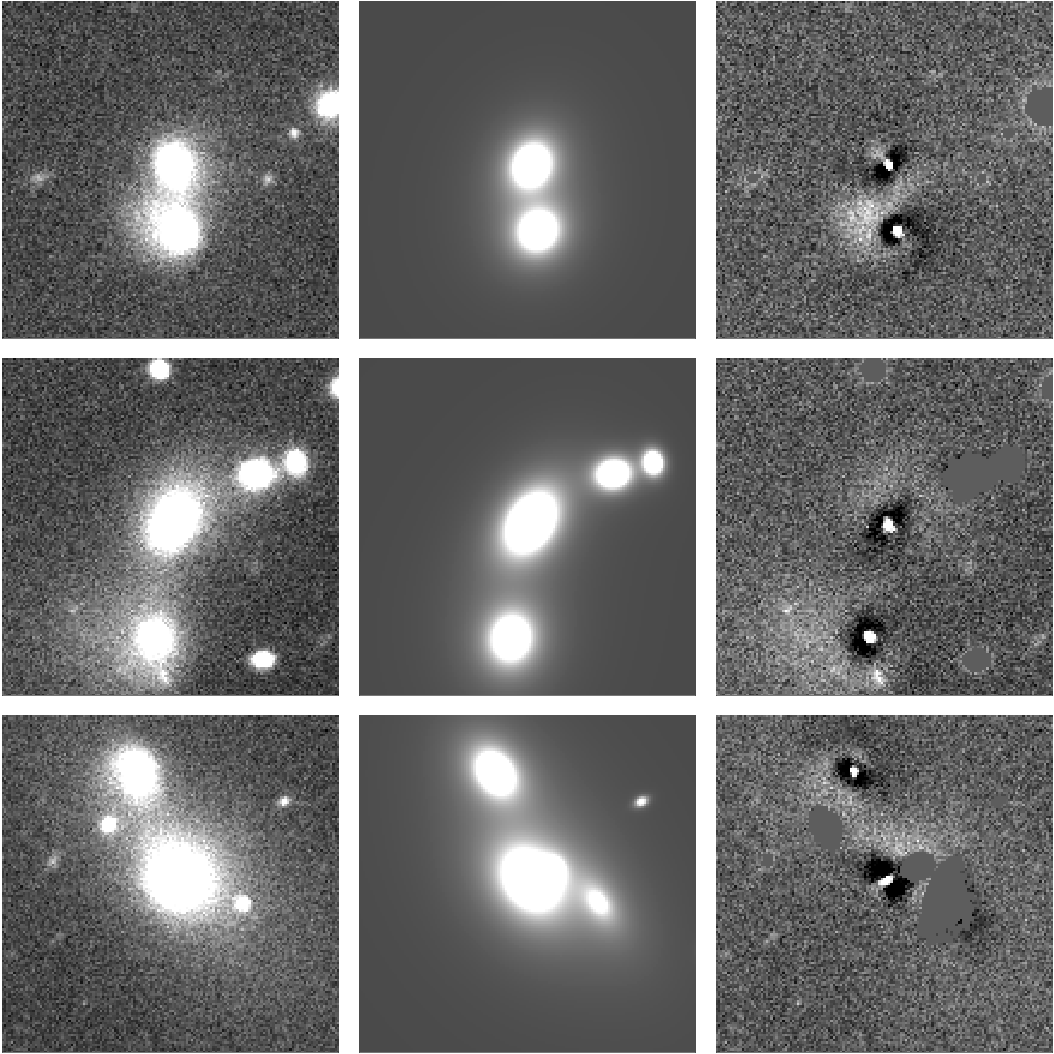}\\[2mm]
\plotone{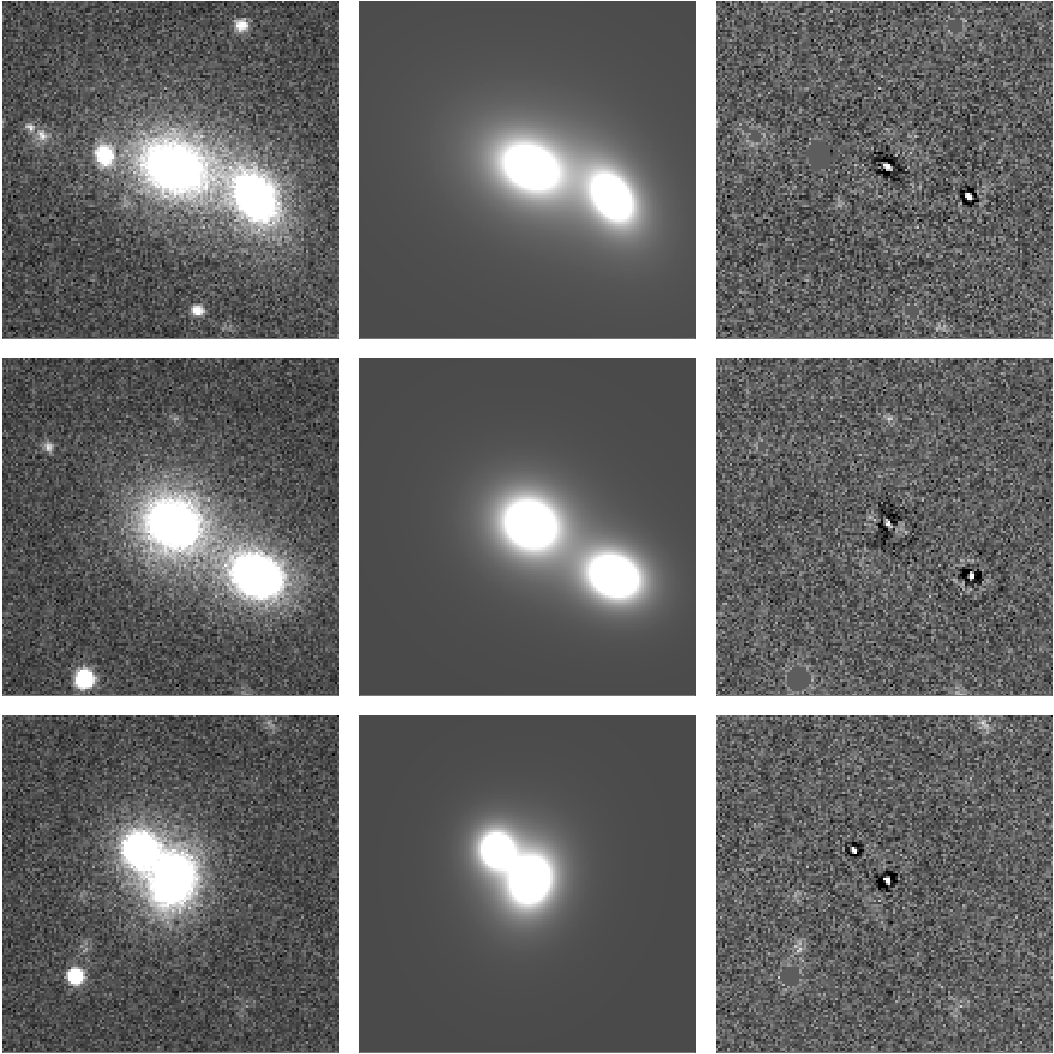}
\caption{Examples of the sky-subtracted images {\it (left)}, the models 
  with GALFIT {\it (middle)}, and the residual images {\it (right)} for
  merging {\it (top 3)} and nonmerging pairs {\it (bottom 3)}.
\label{fitsample1}}
\end{figure}

Figure~\ref{fitsample1} illustrates six examples of GALFIT
fitting. The top three pairs with significant interaction signatures
(e.g., short tidal tails, bridges) in the residual images ({\it right
  panels}) are considered as the merging systems. In contrast, there are no
obvious interaction features in the residual images of the bottom three
examples. The lack of clear interaction signatures suggests that these
pairs may be either due to projection effects or at early stages of
interaction.

\begin{figure}
\epsscale{1.1}
\plotone{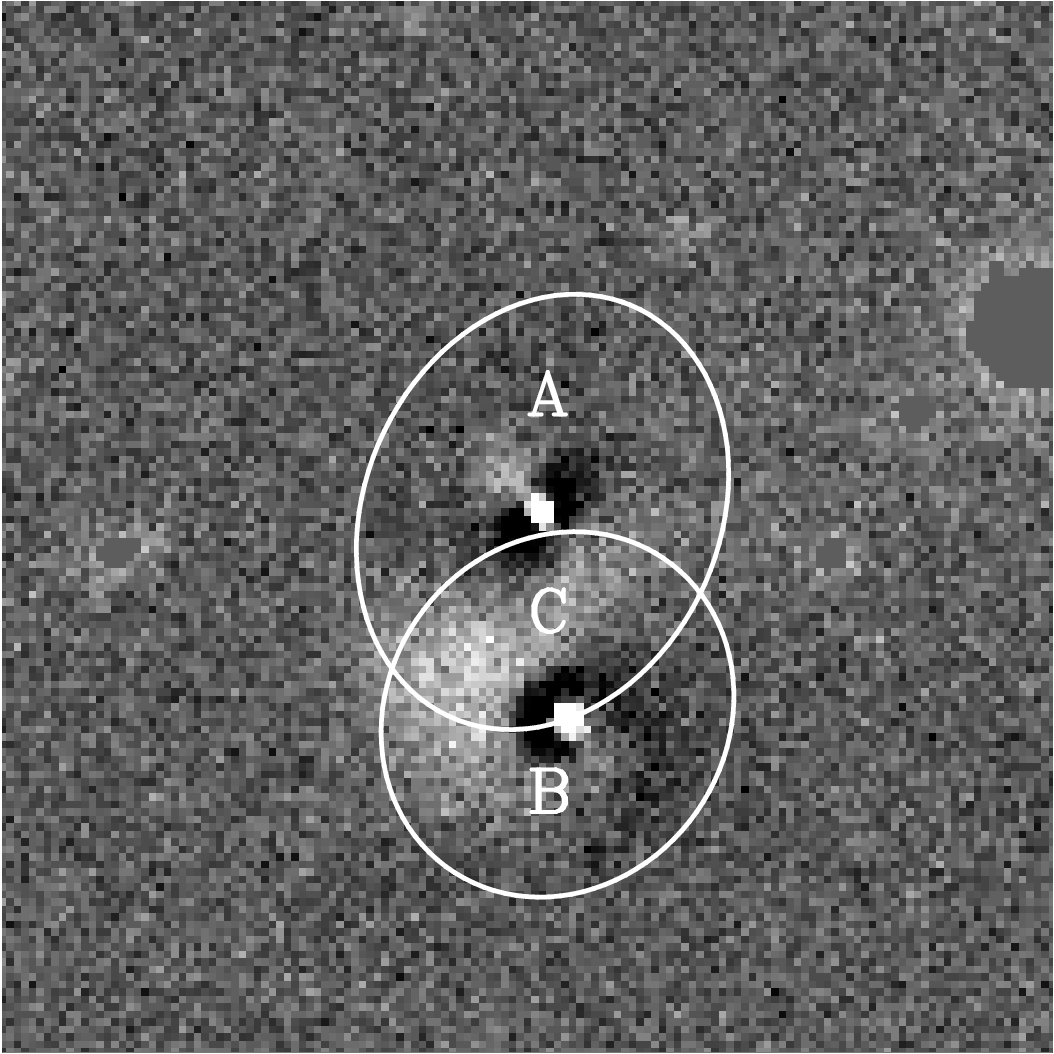}
\caption{An example of the residual image of a galaxy pair. The
ellipses mark the regions with three times of the fitted major and
minor axes of the S{\'e}rsic function for each galaxy. 
The paired galaxies are overlapped in the region of $C$.
\label{illuasy}}
\end{figure}

Quantitative criteria are needed to identify interaction signatures.
We noted that \citet{cbj00} developed a method to estimate the
rotational asymmetry of galaxies and identify major mergers
\citep{cbd+03,dcl+07}. In this paper, we measured the asymmetry
factors of early-type galaxy pairs from the residual images (see an
example in Figure~\ref{illuasy}). The photometric region $A$ or $B$ is
within 3$R_e$ but not overlapped. Here $R_e$ is the effective radius,
a parameter in the fitted S{\'e}rsic function. The overlapped region
is marked as region $C$. Our asymmetry calculation is to measure
the difference between any pixels and those symmetric pixels with
respect to galactic centers. Two different cases exist. One is a pair
of pixels within the region $A$ or $B$; the other is a pixel triplet,
with one pixel in the region $C$, but it has symmetric pixels in
region $A$ and $B$ (or even $C$). It is supposed that there are $N_A$
pairs in region $A$, $N_B$ pairs in region $B$, and $N_C$
pixel triplets, and that the rms value of the residual image is
$\sigma$. We defined the sum of difference squares for three regions
as being
\begin{eqnarray}
\Delta_A&=&\sum^{N_A}[I_A(i)-I_A(i')]^2-2N_A\sigma^2,\nonumber\\
\Delta_B&=&\sum^{N_B}[I_B(i)-I_B(i')]^2-2N_B\sigma^2,\nonumber\\
\Delta_C&=&\sum^{N_C}[I_C(i)-I_A(i')-I_B(i')]^2-3N_C\sigma^2.\nonumber\\
\end{eqnarray}
Here, $I_A(i)-I_A(i')$ is the residual image difference between a
symmetric pixel pair, $i$ and $i'$, in the region $A$;
$I_B(i)-I_B(i')$ is that in the region $B$; and
$I_C(i)-I_A(i')-I_B(i')$ is that for a pixel in region $C$ and its
symmetric pixels in regions $A$ and $B$ (or even $C$). It is 
necessary to subtract the noise power. We also defined
\begin{eqnarray}
S_A&=&\sum^{N_A}[I_A(i)+I_A(i')]^2-2N_A\sigma^2,\nonumber\\
S_B&=&\sum^{N_B}[I_B(i)+I_B(i')]^2-2N_B\sigma^2,\nonumber\\
S_C&=&\sum^{N_C}[I_C(i)+I_A(i')+I_B(i')]^2-3N_C\sigma^2,\nonumber\\
\end{eqnarray}
for normalization. Then, the asymmetry factor, $a$, is defined as
\begin{equation}
a=\frac{\Delta_A+\Delta_B+\Delta_C}{S_A+S_B+S_C}.
\end{equation}

Ideally, $a \sim 0$ is for a galaxy pair without any interaction
feature. A large $a$ means a stronger asymmetric interaction.

\begin{figure}
\epsscale{1.1}
\plotone{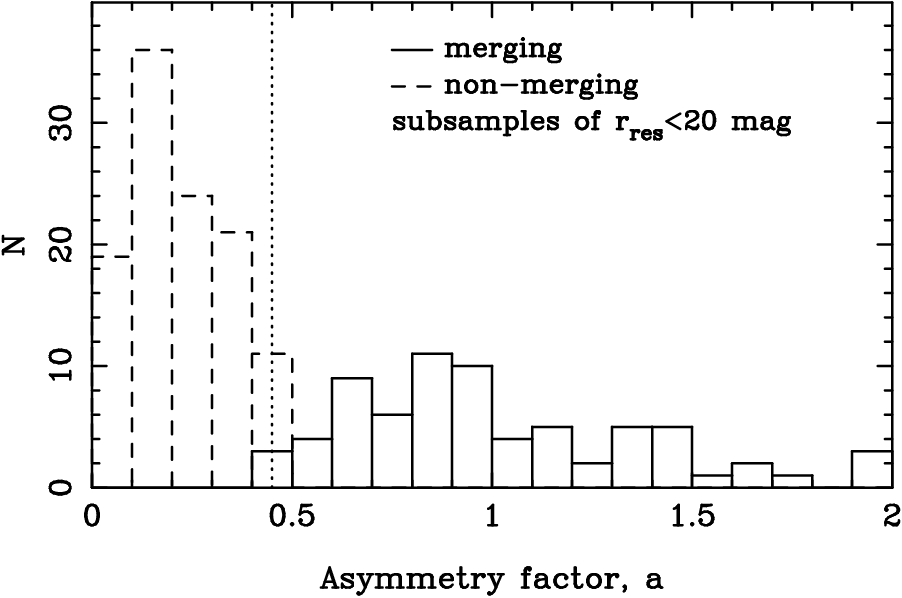}
\caption{Distribution of the asymmetry factors, $a$, for pairs with 
and without clear interaction features.
\label{asymmetry}}
\end{figure}

By visual inspection on the residual images of all projected pairs, we
found 74 pairs with obvious interaction features and 111 pairs clearly
without any feature. The magnitude of residual image $r_{\rm
  res}<20$~mag and the asymmetry factor $a>0.45$ can clearly separate
the two kinds of pairs, as shown in Figure~\ref{asymmetry}. We then
took these criteria to automatically identify the merging pairs from
other projected pairs. Note that the asymmetry factor $a$ sometimes
becomes abnormal when the image of a pair is contaminated by the
objects located within $2R_{90}$ of target galaxies. We verified the
interaction pairs by a further visual check of such a
contamination. Finally, 249 pairs have been identified as merging
pairs, which is about 21\% of all projected pairs.

\subsection{Pair fraction and merger rate}

Our pair sample can be used to estimate the physical pair fraction and
merger rate of luminous early-type galaxies (and SMBHs).

\begin{figure}[tb]
\epsscale{1.1}
\plotone{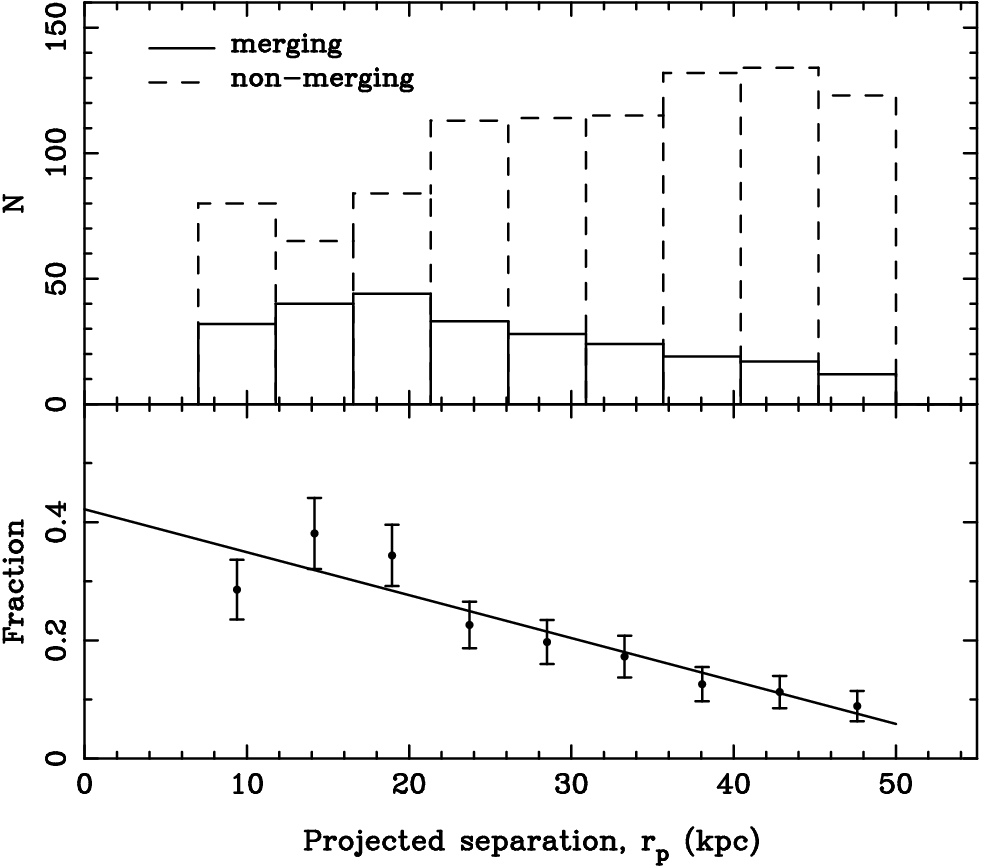}\\[2mm]
\plotone{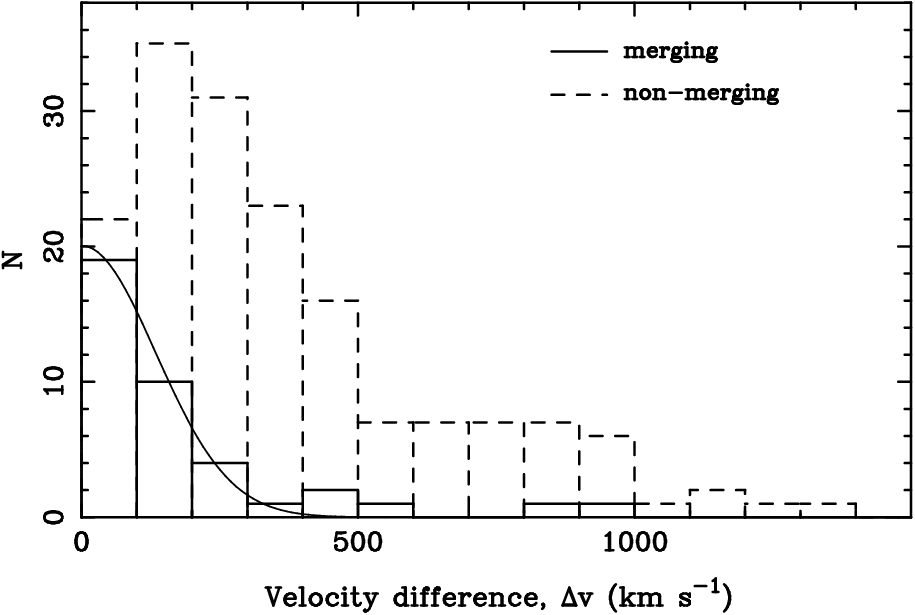}
\caption{Top panel: distribution of the projected separations,
  $r_p$, for the merging and nonmerging pairs and the fraction of
  merging pairs among the projected pairs in each separation bin. {\it
    Lower panel}: distribution of velocity differences $\Delta v$ for
  the merging and nonmerging pairs in the spec--spec sample. For the
  mergers, the velocity dispersions have a Gaussian distribution with
  a standard deviation about 134 ${\rm km~s}^{-1}$.
\label{sepatation}}
\end{figure}

As shown in Figure~\ref{sepatation}, only about 30\%--40\% of projected
pairs with $7<r_p<20$ kpc are mergers. The fraction of mergers among
the projected pairs decreases with the projected separation, to $\sim
20\%$ at 30 kpc and 10\% at 50 kpc. The fraction can be 
formulated to be $(0.42\pm0.04)-(0.007\pm0.001) r_p/{\rm kpc}$. {\it
  Previous pair statistics on the merger fraction with the projected
  pairs should be scaled by these factors.}

Note that there still are galaxy mergers with $r_p<7$~kpc which
likely merge in a shorter timescale, but are not included in our
sample. Taking the number in each separation bin at $r_p<7$~kpc to be
the average number between 7 and 20 kpc, we extrapolated another 57
merging pairs with $r_p<7$~kpc. In total, there should be 306 mergers.

We also checked the distribution of velocity differences for the 
spec--spec pairs. As shown in Figure~\ref{sepatation}, the
velocity differences for merging pairs roughly follow a Gaussian
distribution with a standard deviation of about 134 ${\rm
  km~s}^{-1}$. Mergers rarely have $\Delta v>500$ ${\rm km~s}^{-1}$.

The pair fraction is defined as the number of galaxies in physical
pairs divided by the total number of galaxies (i.e., the number of
physical pairs $\times2$ /the total number of galaxies), which is the
same as $N_c$ in Section 1.1 if there is no triple system. Including the
estimated merging pairs with $r_p<7$ kpc, we found that {\it the
  fraction of galaxies in the merging pairs is 0.6\% or 0.8\%} if one
sets $r_p<30$ or $r_p<50$ kpc, respectively. Note that some physical
bound pairs are not at the stage of merging and may not show strong
interaction signatures; the above-estimated fraction of merging
galaxies should be considered as the lower limit of the physical pair
fraction.

Previous statistics of close pairs usually utilized $\Delta v < 500$
${\rm km~s}^{-1}$ as the criteria for sample selection. If we blindly
used the projected pairs with $\Delta v < 500$ ${\rm km~s}^{-1}$ (see
Figure~\ref{velhist}), without identification of merging features, we
then got the pair fraction, $N_c$, is (1.1$\pm$0.1)\% and
(2.2$\pm$0.2)\% for $7<r_p<30$ kpc and $7<r_p<50$ kpc, respectively,
or (1.0$\pm$0.1)\% for $5<r_p<20$ $h^{-1}$ kpc. The pair fraction is two 
or three time larger than that determined with identification of merging
features, but is consistent with that given by \citet{kss+07}.

\begin{figure}
\epsscale{1.1}
\plotone{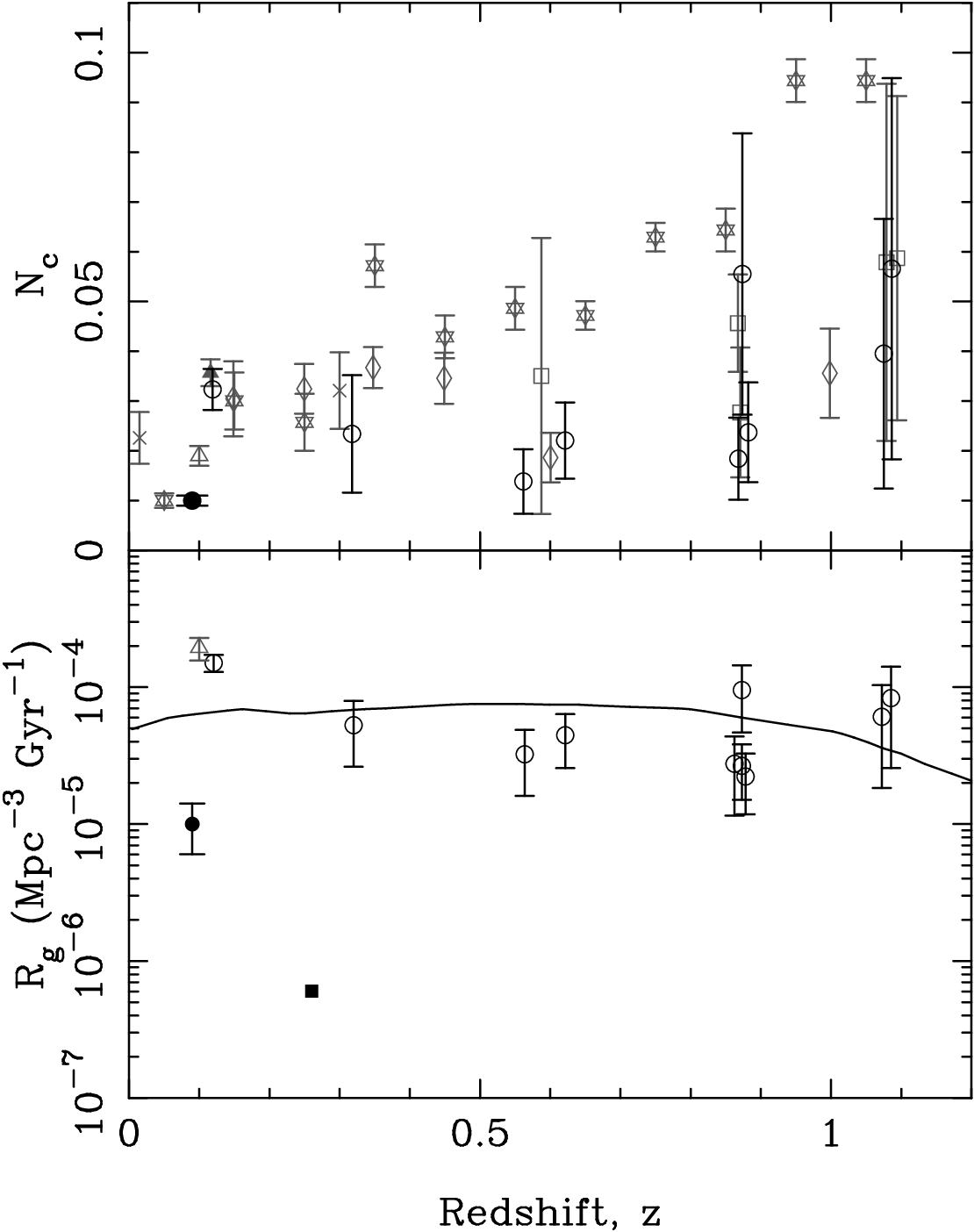}
\caption{The pair fraction we obtained is indicated as the black dot
  at $z=0.09$ and compared with previous statistics of close pairs
  ({\it in the upper panel}): gray crosses indicate the values at
  $z=0.01$ from \citet{pcm+00} and at $z=0.3$ from \citet{ppc+02},
  filled gray triangle at $z=0.12$ from \citet{dld+05}, open gray
  triangle at $z=0.1$ from \citet{dcl+07}, gray diamonds from
  \citet{ccp+00}, gray stars from \citet{kss+07}, black circles from
  \citet{lpk+08}. The values for early-type galaxies are marked as
  black symbols.
The merger rate we determined is indicated as the black dot {\it in
  the lower panel}, and compared with that from \citet{mhc+06} by
black square and \citet{lpk+08} by the black circles for dry merger, and
\citet{dcl+07} by gray triangle for general mergers. The solid line is
the evolution of the merger rate calculated for dry mergers
by \citet{ks08} according to the semianalytical model of galaxy
formation.}
\label{fra_com}
\end{figure}

In Figure~\ref{fra_com}, we have plotted all results on $N_c$ against
redshifts found from the literature by statistics of close pairs. All
values have been normalized to the selection criteria, $5 < r_p < 20$
$h^{-1}$~kpc and $\Delta v < 500$ ${\rm km~s}^{-1}$
\citep{pcm+00}. Our result is plotted as a black dot at the mean
redshift $\langle z \rangle=0.09$.  The gray cross at redshift
$z=0.01$ is taken from \citet{pcm+00}, and that at redshift $z=0.3$
is from \citet{ppc+02}; the filled gray triangle at redshift $z=0.12$
is from \citet{dld+05}, and the open gray triangle at redshift $z=0.1$ is from
\citet{dcl+07}.  \citet{ccp+00} derived fractions with $r_p\le 50$
$h^{-1}$~kpc and $\Delta v<1000$ ${\rm km~s}^{-1}$. They found that
the number of pairs with $r_p\le 50$ $h^{-1}$~kpc is 3.8 times that
with $r_p\le 20$ $h^{-1}$~kpc. The ratio between the number of pairs
with $\Delta v<500$ ${\rm km~s}^{-1}$ and that with $\Delta v<1000$
${\rm km~s}^{-1}$ is 0.9. We normalized the values and showed them as
gray diamonds.  \citet{kss+07} gave the pair fractions with 5 $<r_p<
20$ kpc adopting $h=0.7$. We corrected their values based on the fact
that the pair fraction is roughly proportional to the maximum
projected separation \citep{pcm+00}. The normalized values after
correction are shown as gray stars. \citet{lkw+04,lpk+08} studied
close pairs with 10 $<r_p< 30$, 50, 100 $h^{-1}$~kpc. The values from 
\citet{lkw+04} are shown as gray squares. \citet{lpk+08} provided
pair fractions for different types of pairs. For a direct comparison
with our result, the black circles in the upper panel of
Figure~\ref{fra_com} are the pair fractions for early-type mergers.
The normalized pair fraction from \citet{lpk+08} is 3.2\% at
redshift $z=0.12$, three times that of this work.

Now we can estimate the merger rate of luminous early-type galaxies
from our sample. The merger rate is defined as the number of merger
events per unit time per comoving volume:
\begin{equation}
R_{\rm g}(z)=n(z)C_{\rm mg}/T_{\rm mg},
\end{equation}
where $T_{\rm mg}$ is the timescale for a pair to merge, $C_{\rm mg}$
is the fraction of pairs that will merge within $T_{\rm mg}$, and
$n(z)$ is the number density of pairs, i.e., the number of pairs 
divided by the comoving volume.
The merging timescale depends on separation, relative velocity, and
mass ratios of galaxies. It has always been assumed that the merging
pairs will coalesce within some binary orbits.  \citet{pcm+00} and
\citet{con06} assumed an average value of 0.5 Gyr for the merging
timescale.  \citet{bnm+06} suggested $T_{\rm
  mg}\sim0.15\pm0.05$ Gyr for massive merging pairs with
$r_p<20$~kpc. \citet{rfv07} found a merging timescale of $\sim0.11$
Gyr for a very luminous merger system, such as CL0958+4702.  Here, we
adopted an average merging timescale of 0.3$^{+0.2}_{-0.1}$ Gyr for
the mergers in our sample, which show the distinct interaction
features. We further assumed $C_{\rm mg}=1$ for our merging pairs, as
suggested by simulations \citep{bnm+06}.  Based on our sample of
merging pairs including 57 pairs with $r_p<7$~kpc, we found that the
number density of merging pairs is $n=3.1\times10^{-6}~{\rm Mpc^{-3}}$
at redshifts $z<0.12$.  Putting these all together, we derived {\it the
  comoving volume merger rate},
\begin{equation}
R_{\rm g}=(1.0\pm0.4)\times10^{-5}~{\rm Mpc^{-3}~Gyr^{-1}}.
\label{rg}
\end{equation}
The uncertainty mainly depends on the merging timescale. 

In the lower panel of Figure~\ref{fra_com}, we made a comparison of
our merger rate with those from the literature. Our merger rate (black
dot) is much larger than $6\times10^{-7}~{\rm Mpc^{-3}~Gyr^{-1}}$
(filled square) by \citet{mhc+06} for the SDSS Luminous Red
Galaxies, but much smaller than $\sim10^{-4}~{\rm Mpc^{-3}~Gyr^{-1}}$
(open circles) by \citet{lpk+08} for early-type galaxy pairs in a
number of surveys and $5.2\times10^{-4}~h^3~{\rm Mpc^{-3}~Gyr^{-1}}$
(gray triangle) by \citet{dcl+07} for general merger from the
Millennium Galaxy Catalog. Based on a semianalytical model of galaxy
formation, \citet{ks08} found that the merger rate of dry mergers with
masses $M>6.3\times 10^{10}$ $M_{\odot}$ is almost a constant value of
$6\times10^{-5}~{\rm Mpc^{-3}~Gyr^{-1}}$ at redshifts $z<0.8$ (see the
line in the lower panel of Figure~\ref{fra_com}).

The discrepancy probably arises from different magnitude limits for
the sample selections. The lower magnitude limit in \citet{lpk+08} is
$M_B=-19$, which appears to be one magnitude fainter than the limit in
our sample after the color correction of $B-r=1.2$ \citep{jsr+05}.
Given $-23.2<M_g<-21.2$, after $k$-corrections to redshift $z=0.3$,
the number density of the SDSS Luminous Red Galaxies in
\citet{mhc+06} is $10^{-4}~{\rm Mpc^{-3}}$, 10\% of that for our
sample and 4\% for the fainter galaxies ($-21<M_B<-19$) by
\citet{dcl+07} and \citet{lpk+08}.  Relating the magnitudes of mergers
to the masses of mergers \citep{jk07}, we scaled up our merger rate
with $M>4.2\times10^{11}~M_{\odot}$ by a factor of 5--10 to the
value with $M>6.3\times10^{10}~M_{\odot}$ according to the cumulative
function of merger rate against the mass of the merger \citep[Figure 2
  of][]{ks08}. In general, our corrected merger rate appears to be in
good agreement with that from the model of \citet{ks08}.

\section{SMBH Mergers in the pairs and Gravitational Wave Background}

In the following, we assume that SMBH binaries can be formed in 
galaxy mergers, and efficiently driven to coalescence with GWs
radiation.
The detailed physical processes of an SMBH binary shrinking to the GW
regime are far from clear \citep{jb03,svc08}. But very few SMBH
binaries have been found in the centers of galaxies
\citep[e.g.,][]{ooi+85,rtz+06}, implying that most of the SMBH binaries
from galaxy merging must have lost enough momentum and merged in a
very short period.
A large number of merging-induced coalescences generate a stochastic
GW background, which could be a promising GW source to be detected 
by using pulsar timing measurement.

\subsection{The masses of SMBHs in mergers}

The mass of an SMBH, $M$, in the center of galaxy is tightly related to
the velocity dispersion, $\sigma$, the luminosity, $L$, or the mass of
bulge $M_{\rm bulge}$, \citep{mtr+98,mf01,md02,mh03,hr04}. The SMBH
mass correlates more tightly with $\sigma$ than $L$
\citep{fm00,gbb+00}. For massive early-type galaxies, however,
whether $L$ or $\sigma$ is a better predictor of the SMBH mass is
still an open question. The $M$--$L$ relation predicts more $10^9$
$M_{\odot}$ SMBHs than the $M$--$\sigma$ relation does
\citep{lfr+07}. Recently, \citet{tbh+07} suggested that the $M$--$\sigma$
relation may not follow a single power law.  Because of the lack of the
velocity dispersions for all pairs, we estimated the masses of SMBHs
in mergers using the $M$--$L$ relation \citep{tbh+07}
\begin{equation}
\log M=(8.69\pm0.10) - \frac{(1.31\pm0.15)}{2.5}(M_r+22).
\end{equation}
As stated in Section 2.2, the magnitudes of paired galaxies are taken from
the model-fitted values in $r$ band.

\begin{figure}
\epsscale{1.1}
\plotone{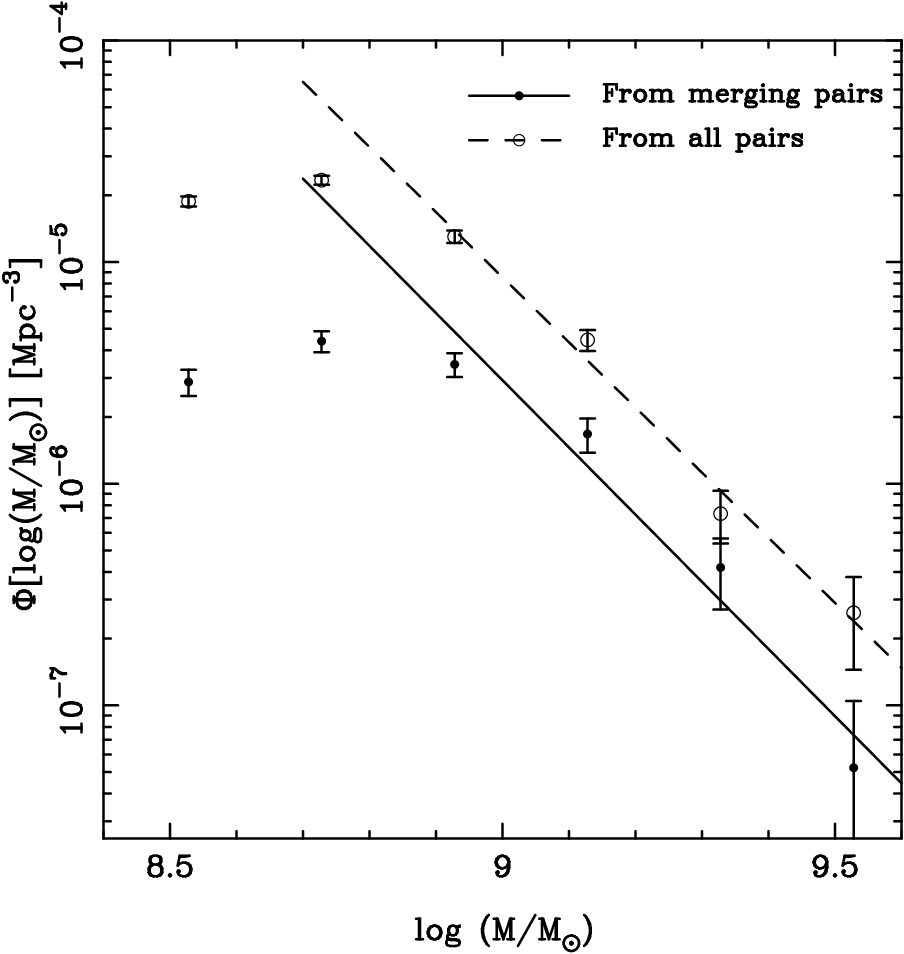}
\caption{The chirp mass distribution of the SMBH binaries in the
  mergers (filled circle) and all projected pairs (open circle). The
  lines represent the best fits of power laws.
\label{chirp}}
\end{figure}

From these masses of SMBHs, we determined the chirp masses of the SMBH
binaries in the mergers (Eq.~\ref{chirpmass}). We found that the chirp
mass distribution of SMBH binaries (Figure~\ref{chirp}) can be fitted
with a power law
\begin{equation}
\Phi[\log (M /M_{\odot})]=(21.7\pm4.2)-(3.0\pm0.5)\log M/M_{\odot}.
\label{chmass}
\end{equation}
The two points of low chirp masses were excluded in the fitting
because they are underestimated due to the sample selection effect. The
magnitude cutoff for galaxies is $M_r=-21.5$, so that pairs with one
galaxy fainter than $M_r=-21.5$ are missing in our sample.

\subsection{The amplitude of gravitational wave strain}

The spectrum of the GW background from the SMBH mergers can be
calculated from the chirp mass distribution and the coalescence rate
of a population of SMBH binaries (see Equation~\ref{hs}--\ref{numbin}). Based
on the newly determined chirp mass distribution and merger rate, we
can estimate the amplitude of the GW strain (see Equation~\ref{timeeq}--\ref{numbin}). 
Recall that (see Section 1.2) no observation-based chirp mass
distributions were ever available in any previous calculations.

Another necessary parameter for the estimation of the GW strain is the
merger rate evolution index, $m$. Previous studies (see Section 1.1) have
parameterized the merger rate of galaxies as a function of redshift in
the form of $(1+z)^m$ with an uncertainty of $m$ in the range of -1 to
3. In the following discussion, we assume that the merger rate of
galaxies is equal to that of SMBHs and both evolve in the form of
$R(z)\propto(1+z)^m$. We also assume that the chirp mass distribution
does not depend on redshift. After putting the local merger rate from
Eq.\ref{rg} and the chirp mass distribution from Eq.\ref{chmass} into
Eq.~\ref{hcf}, we obtained 
\begin{equation}
h_c(f)=1.1\times 10^{-15}\Big(\frac{f}{\rm yr^{-1}}\Big)^{-2/3}I^{1/2},
\label{hcfnew}
\end{equation}
in the frequency range of $f\sim 10^{-9}$ to $10^{-7}$ Hz. Here 
\begin{equation}
I=\int \frac{dz}{E(z)(1+z)^{4/3-m}}.
\end{equation}
The integration between redshift $z=0$ and 3 can be simplified as
\begin{equation}
I^{1/2}\approx (0.2m+0.56)^{3.6}+0.7,
\end{equation}
in the range of $m=-1$ to 4.

For the calculation of $h_c(f)$ above, we considered only the massive
mergers. Lower mass systems (i.e., $M_r>-21.5$) is found to add only
1\% of the strain amplitude, if the SMBH population at lower mass 
in \citet{bdf+07} and the merging fraction of this work are considered.

\begin{figure}
\epsscale{1.1}
\plotone{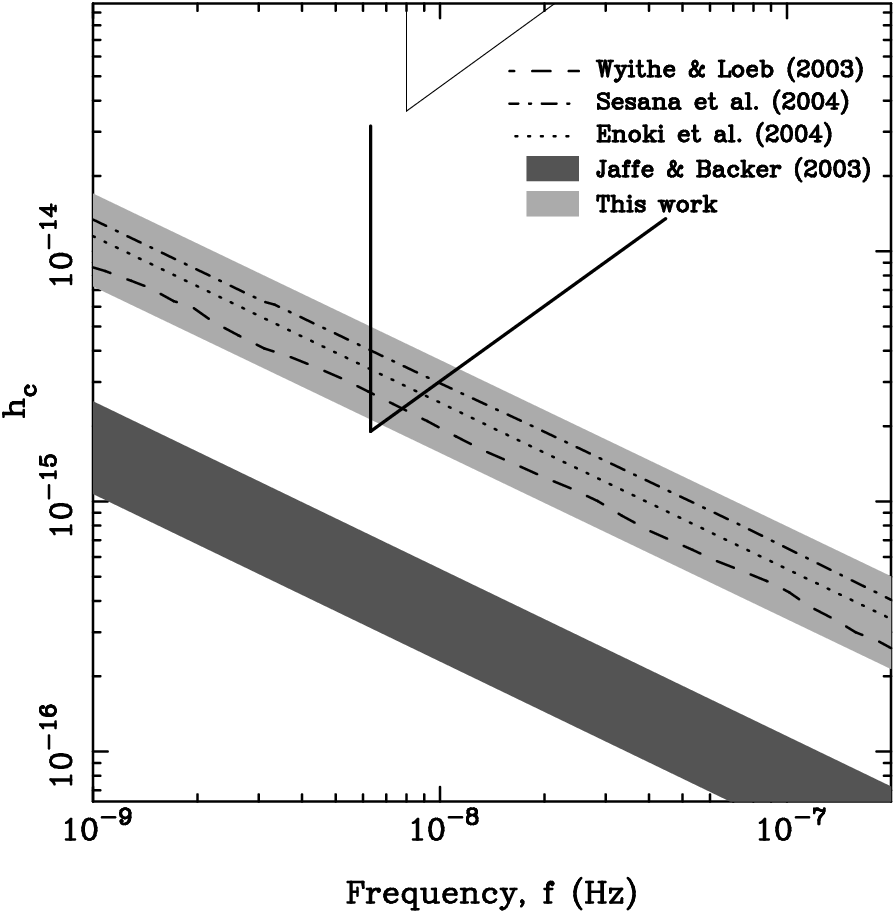}
\caption{Strain spectrum of the stochastic GW background calculated in
  the range of $m=-1$ to 2 using Eq.\ref{hcfnew} (light gray
  area). The strain spectrum obtained by \citet{jb03} is shown as the
  dark gray area, which was calculated with a different merger rate 
  and SMBH population. The strain spectrum obtained by \citet{wl03} is
  shown as the dashed line, by \citet{shmv04} as dot-dashed line, and
  by \citet{eins04} as the dotted line. The upper thin solid and lower
  thick solid lines represent the upper limit of GWs constrained by
  using available pulsar-timing data sets and simulated data sets of
  the complete PPTA \citep{jhv+06}.
\label{strain}}
\end{figure}

We showed our strain spectrum together with previous estimations in
Figure~\ref{strain}. \citet{wl03} calculated the strain spectrum of
the GW background with a theoretical model, giving the strain of
(0.5--0.9)$\times10^{-15}(f/{\rm yr}^{-1})^{-2/3}$ for different
initial parameters of the model (see their Figure 6).  The amplitude
corresponds to the evolution index $m$ in the range of $-3.6$ to 0
according to the local merger rate and the SMBH population in this
work. The amplitudes obtained by \citet{shmv04} and \citet{eins04}
correspond to the indices $m=1.5$ and $m=0.9$, respectively.
\citet{jb03} quoted the pair fraction from \citet{ccp+00} and the SMBH
population transformed from \citet{fs91}. They showed that the GW
background is dominated by the systems of $\sim
10^8$~$M_{\odot}$. Thus, the GW strain amplitude they estimated is a
few times lower than this work and others \citep{wl03,shmv04,eins04}.

The strain amplitude we estimated is comparable to the GW detection
sensitivity of the pulsar timing project.
Using currently available pulsar timing data sets, one can place the upper
limit of the GW background strain or the GW energy density.  The GW energy
density per unit logarithmic frequency interval can be written as
\citep{phi01}
\begin{equation}
\Omega(f)=\frac{2\pi^2}{3H_0^2}f^2h_c(f)^2.
\end{equation}
Using the eight year timing data of PSR B1855+09, \citet{ktr94} obtained
$\Omega h^2<6\times10^{-8}$ for the cosmic GW background. Using the 17
year data of this object, \citet{lom02} got
$\Omega(f)h^2<2\times10^{-9}$.  \citet{jhv+06} developed a new
technique to place better limits using the available pulsar-timing
data sets in the Parks pulsar-timing array (PPTA) project
\citep{hob05,man06} and the simulation data sets (20 pulsars with rms
timing residual of 100~ns over five years) of the future complete
PPTA. They derived the upper limits of the GW background emitted from
the SMBH coalescence to be $1.1\times 10^{-14}(f/{\rm
  yr}^{-1})^{-2/3}$ and $6.5\times 10^{-16}(f/{\rm yr}^{-1})^{-2/3}$
for the available data of the PPTA and the complete PPTA,
respectively.
The upper limit from the available pulsar-timing data corresponds to
an index $m=5.8$, which is not very meaningful from our knowledge of 
the evolution of the galaxy merger rate. The complete PPTA data in future 
could limit the index $m$ to $-1.6$.

\section{Conclusions}

Previously almost all calculations for GW background have been derived from
the theoretical models of galaxy mergers and simulations, with the
merger rate or/and SMBH mass function ``calibrated'' by the observed
values.
\citet{jb03} have directly used the observed galaxy merger rate and
the SMBH mass function for this purpose.
However, when deriving the merger rate from projected galaxy pairs, very
few previous authors have carefully checked the merging fraction of a
large sample of pairs. Normally, the merging conditions were set as
projected separations $r_p<20$ $h^{-1}$~kpc and the radial velocity
differences $\Delta v<500$ ${\rm km~s}^{-1}$, which are certainly
necessary but not sufficient conditions.

We have made a careful selection of a large complete volume-limited
sample of projected close pairs ($r_p<50$ kpc) of luminous early-type
galaxies ($M_r<-21.5$) in the local universe ($z<0.12$) from the SDSS
photometric and spectroscopic data. 71\% of the pairs have $\Delta v<500$
${\rm km~s}^{-1}$, and 21\% of the pairs show merging
features. Considering the total number of all luminous early-type
galaxies, we found that 0.8\% of the galaxies are merging.
From the merging pairs, we derived a comoving volume merger rate of
$\sim (1.0\pm0.4)\times 10^{-5}$ ${\rm Mpc^{-3}~Gyr^{-1}}$ for
luminous early-type galaxies. This is a direct observational
determination of the merger rate of luminous galaxies in the local
universe. Our merger rate is larger than that derived from the SDSS
Luminous Red Galaxies \citep{mhc+06}.

For the first time, from the identified merging pairs of a complete
sample of luminous early-type galaxies, we found that the chirp mass
distribution of SMBH binaries can be described by a power law. With
less assumptions than previous authors, we obtained the strain
amplitude of the GW background from coalescence of SMBH binaries in
frequency range $10^{-9}$--$10^{-7}$~Hz, $h_c(f)\sim
10^{-15}(f/{\rm yr}^{-1})^{-2/3}$. The uncertainty of the GW
background estimation now mainly comes from poor knowledge on the
merger rate evolution, the SMBH population and unknown processes for an
SMBH binary to be driven into the GW regime.

\acknowledgments

We thank the anonymous referee, Professor Fredrick A. Jenet and Dr. R.N.Manchester for
helpful comments. We thank Dr. JunHui Zhao and Dr. XiaoHui Sun for
careful reading of the manuscript.
This work is partially supported by the National Natural Science
Foundation (NNSF) of China (10521001, 10773016) and the National Key
Basic Research Science Foundation of China (2007CB815403) and the
Doctoral Foundation of SYNU of China (054-55440105020).
Funding for the SDSS and SDSS-II has been provided by the Alfred
P. Sloan Foundation, the Participating Institutions, the National
Science Foundation, the U.S. Department of Energy, the National
Aeronautics and Space Administration, the Japanese Monbukagakusho, the
Max Planck Society, and the Higher Education Funding Council for
England. The SDSS Web site is http://www.sdss.org/.
The SDSS is managed by the Astrophysical Research Consortium for the
Participating Institutions. The Participating Institutions are the
American Museum of Natural History, Astrophysical Institute Potsdam,
University of Basel, Cambridge University, Case Western Reserve
University, University of Chicago, Drexel University, Fermilab, the
Institute for Advanced Study, the Japan Participation Group, Johns
Hopkins University, the Joint Institute for Nuclear Astrophysics, the
Kavli Institute for Particle Astrophysics and Cosmology, the Korean
Scientist Group, the Chinese Academy of Sciences (LAMOST), Los Alamos
National Laboratory, the Max Planck Institute for Astronomy (MPIA),
the Max Planck Institute for Astrophysics (MPA), New Mexico State
University, Ohio State University, University of Pittsburgh,
University of Portsmouth, Princeton University, the United States
Naval Observatory, and the University of Washington.

\bibliographystyle{apj}
\bibliography{journals,gw}

\begin{thebibliography}{78}
\expandafter\ifx\csname natexlab\endcsname\relax\def\natexlab#1{#1}\fi

\bibitem[{Abramovici {et~al.}(1992)Abramovici et al.}]{aad+92}
{Abramovici}, A. et al. 1992, Science, 256, 325

\bibitem[{Adelman-McCarthy {et~al.}(2008)Adelman-McCarthy et al.}]{dr6}
{Adelman-McCarthy}, J.~K. et al. 2008, \apjs, 175, 297

\bibitem[{{Armitage} \& {Natarajan}(2002)}]{an02}
{Armitage}, P.~J. \& {Natarajan}, P. 2002, \apjl, 567, L9

\bibitem[{{Begelman} {et~al.}(1980){Begelman}, {Blandford}, \& {Rees}}]{bbr80}
{Begelman}, M.~C., {Blandford}, R.~D., \& {Rees}, M.~J. 1980, \nat, 287, 307

\bibitem[{Bell {et~al.}(2006{\natexlab{a}})Bell et al.}]{bnm+06}
{Bell}, E.~F. et al. 2006{\natexlab{a}}, \apj, 640, 241

\bibitem[{Bell {et~al.}(2006{\natexlab{b}})Bell et al.}]{bps+06}
{Bell}, E.~F. et al. 2006{\natexlab{b}}, \apj, 652, 270

\bibitem[{{Benson} {et~al.}(2007){Benson}, {D{\v z}anovi{\'c}}, {Frenk}, \&
  {Sharples}}]{bdf+07}
{Benson}, A.~J., {D{\v z}anovi{\'c}}, D., {Frenk}, C.~S., \& {Sharples}, R.
  2007, \mnras, 379, 841

\bibitem[{{Blanton} {et~al.}(2003){Blanton}, {Lin}, {Lupton}, {Maley}, {Young},
  {Zehavi}, \& {Loveday}}]{bll+03}
{Blanton}, M.~R., {Lin}, H., {Lupton}, R.~H., {Maley}, F.~M., {Young}, N.,
  {Zehavi}, I., \& {Loveday}, J. 2003, \aj, 125, 2276

\bibitem[{{Blanton} \& {Roweis}(2007)}]{br07}
{Blanton}, M.~R. \& {Roweis}, S. 2007, \aj, 133, 734

\bibitem[{{Bundy} {et~al.}(2004){Bundy}, {Fukugita}, {Ellis}, {Kodama}, \&
  {Conselice}}]{bfe+04}
{Bundy}, K., {Fukugita}, M., {Ellis}, R.~S., {Kodama}, T., \& {Conselice},
  C.~J. 2004, \apjl, 601, L123

\bibitem[{{Burkey} {et~al.}(1994){Burkey}, {Keel}, {Windhorst}, \&
  {Franklin}}]{bkw+94}
{Burkey}, J.~M., {Keel}, W.~C., {Windhorst}, R.~A., \& {Franklin}, B.~E. 1994,
  \apjl, 429, L13

\bibitem[{{Carlberg} {et~al.}(1994){Carlberg}, {Pritchet}, \&
  {Infante}}]{cpi94}
{Carlberg}, R.~G., {Pritchet}, C.~J., \& {Infante}, L. 1994, \apj, 435, 540

\bibitem[{Carlberg {et~al.}(2000)Carlberg et al.}]{ccp+00}
{Carlberg}, R.~G. et al. 2000, \apjl, 532, L1

\bibitem[{{Conselice}(2006)}]{con06}
{Conselice}, C.~J. 2006, \apj, 638, 686

\bibitem[{{Conselice} {et~al.}(2003{\natexlab{a}}){Conselice}, {Bershady},
  {Dickinson}, \& {Papovich}}]{cbd+03}
{Conselice}, C.~J., {Bershady}, M.~A., {Dickinson}, M., \& {Papovich}, C.
  2003{\natexlab{a}}, \aj, 126, 1183

\bibitem[{{Conselice} {et~al.}(2000){Conselice}, {Bershady}, \&
  {Jangren}}]{cbj00}
{Conselice}, C.~J., {Bershady}, M.~A., \& {Jangren}, A. 2000, \apj, 529, 886

\bibitem[{{Conselice} {et~al.}(2003{\natexlab{b}}){Conselice}, {Chapman}, \&
  {Windhorst}}]{ccw03}
{Conselice}, C.~J., {Chapman}, S.~C., \& {Windhorst}, R.~A. 2003{\natexlab{b}},
  \apjl, 596, L5

%\bibitem[{{Damour} \& {Vilenkin}(2005)}]{dv05}
%{Damour}, T. \& {Vilenkin}, A. 2005, \prd, 71, 063510

\bibitem[{{De Propris} {et~al.}(2007){De Propris}, {Conselice}, {Liske},
  {Driver}, {Patton}, {Graham}, \& {Allen}}]{dcl+07}
{De Propris}, R., {Conselice}, C.~J., {Liske}, J., {Driver}, S.~P., {Patton},
  D.~R., {Graham}, A.~W., \& {Allen}, P.~D. 2007, \apj, 666, 212

\bibitem[{{De Propris} {et~al.}(2005){De Propris}, {Liske}, {Driver}, {Allen},
  \& {Cross}}]{dld+05}
{De Propris}, R., {Liske}, J., {Driver}, S.~P., {Allen}, P.~D., \& {Cross},
  N.~J.~G. 2005, \aj, 130, 1516

\bibitem[{{Detweiler}(1979)}]{det79}
{Detweiler}, S. 1979, \apj, 234, 1100

\bibitem[{{Dotti} {et~al.}(2007){Dotti}, {Colpi}, {Haardt}, \&
  {Mayer}}]{dch+07}
{Dotti}, M., {Colpi}, M., {Haardt}, F., \& {Mayer}, L. 2007, \mnras, 379, 956

\bibitem[{{Enoki} {et~al.}(2004){Enoki}, {Inoue}, {Nagashima}, \&
  {Sugiyama}}]{eins04}
{Enoki}, M., {Inoue}, K.~T., {Nagashima}, M., \& {Sugiyama}, N. 2004, \apj,
  615, 19

\bibitem[{{Escala} {et~al.}(2004){Escala}, {Larson}, {Coppi}, \&
  {Mardones}}]{elc+04}
{Escala}, A., {Larson}, R.~B., {Coppi}, P.~S., \& {Mardones}, D. 2004, \apj,
  607, 765

%\bibitem[{Fan {et~al.}(2001)Fan et al.}]{fnl+01}
%{Fan}, X. et al. 2001, \aj, 122, 2833

\bibitem[{{Ferguson} \& {Sandage}(1991)}]{fs91}
{Ferguson}, H.~C. \& {Sandage}, A. 1991, \aj, 101, 765

\bibitem[{{Ferrarese} \& {Merritt}(2000)}]{fm00}
{Ferrarese}, L. \& {Merritt}, D. 2000, \apjl, 539, L9

\bibitem[{{Gebhardt} {et~al.}(2000){Gebhardt} et al.}]{gbb+00}
{Gebhardt}, K. et al. 2000, \apjl, 539, L13

\bibitem[{{Gould} \& {Rix}(2000)}]{gr00}
{Gould}, A. \& {Rix}, H.-W. 2000, \apjl, 532, L29

\bibitem[{{Haehnelt}(1994)}]{hae94}
{Haehnelt}, M.~G. 1994, \mnras, 269, 199

\bibitem[{{H{\"a}ring} \& {Rix}(2004)}]{hr04}
{H{\"a}ring}, N. \& {Rix}, H.-W. 2004, \apjl, 604, L89

\bibitem[{{Hobbs}(2005)}]{hob05}
{Hobbs}, G. 2005, PASA, 22, 179

\bibitem[{{Hughes}(2002)}]{hug02}
{Hughes}, S.~A. 2002, \mnras, 331, 805

\bibitem[{{Jaffe} \& {Backer}(2003)}]{jb03}
{Jaffe}, A.~H. \& {Backer}, D.~C. 2003, \apj, 583, 616

\bibitem[{{Jenet} {et~al.}(2005){Jenet}, {Hobbs}, {Lee}, \&
  {Manchester}}]{jhl+05}
{Jenet}, F.~A., {Hobbs}, G.~B., {Lee}, K.~J., \& {Manchester}, R.~N. 2005,
  \apjl, 625, L123

\bibitem[{Jenet {et~al.}(2006)Jenet et al.}]{jhv+06}
{Jenet}, F.~A. et al. 2006, \apj, 653, 1571

\bibitem[{Jester {et~al.}(2005)Jester et al.}]{jsr+05}
{Jester}, S. et al. 2005, \aj, 130, 873

\bibitem[{{Jiang} \& {Kochanek}(2007)}]{jk07}
{Jiang}, G. \& {Kochanek}, C.~S. 2007, \apj, 671, 1568

\bibitem[{Kampczyk {et~al.}(2007)Kampczyk et al.}]{klc+07}
{Kampczyk}, P. et al. 2007, \apjs, 172, 329

\bibitem[{Kartaltepe {et~al.}(2007)Kartaltepe et al.}]{kss+07}
{Kartaltepe}, J.~S. et al. 2007, \apjs, 172, 320

\bibitem[{{Kaspi} {et~al.}(1994){Kaspi}, {Taylor}, \& {Ryba}}]{ktr94}
{Kaspi}, V.~M., {Taylor}, J.~H., \& {Ryba}, M.~F. 1994, \apj, 428, 713

\bibitem[{{Khochfar} \& {Silk}(2008)}]{ks08}
{Khochfar}, S. \& {Silk}, J. 2008, arXiv:astro-ph/0809.1734

%\bibitem[{{Kormendy} \& {Richstone}(1992)}]{kr92}
%{Kormendy}, J. \& {Richstone}, D. 1992, \apj, 393, 559

\bibitem[{Lauer {et~al.}(2007)Lauer et al.}]{lfr+07}
{Lauer}, T.~R. et al. 2007, \apj, 662, 808

\bibitem[{{Lavery} {et~al.}(2004){Lavery}, {Remijan}, {Charmandaris}, {Hayes},
  \& {Ring}}]{lrc+04}
{Lavery}, R.~J., {Remijan}, A., {Charmandaris}, V., {Hayes}, R.~D., \& {Ring},
  A.~A. 2004, \apj, 612, 679

\bibitem[{Le F{\`e}vre {et~al.}(2000)Le F{\`e}vre et al.}]{lal+00}
{Le F{\`e}vre}, O. et al. 2000, \mnras, 311, 565

\bibitem[{Lin {et~al.}(2004)Lin et al.}]{lkw+04}
{Lin}, L. et al. 2004, \apjl, 617, L9

\bibitem[{Lin {et~al.}(2008)Lin et al.}]{lpk+08}
{Lin}, L., et al. 2008, \apj, 681, 232

\bibitem[{{Liu} {et~al.}(2008){Liu}, {Xia}, {Mao}, {Wu}, \& {Deng}}]{lxm+08}
{Liu}, F.~S., {Xia}, X.~Y., {Mao}, S., {Wu}, H., \& {Deng}, Z.~G. 2008, \mnras,
  385, 23

\bibitem[{{Lommen}(2002)}]{lom02}
{Lommen}, A.~N. 2002, in WE-Heraeus Seminar on Neutron Stars, Pulsars, and Supernova Remnants, ed.
  W.~{Becker}, H.~{Lesch}, \& J.~{Tr{\"u}mper} (Garching: MPE), 114

\bibitem[{Lotz {et~al.}(2008)Lotz et al.}]{ldf+08}
{Lotz}, J.~M. et al. 2008, \apj, 672, 177

\bibitem[{{Maggiore}(2000)}]{mag00}
{Maggiore}, M. 2000, \physrep, 331, 283

\bibitem[{Magorrian {et~al.}(1998)Magorrian et al.}]{mtr+98}
{Magorrian}, J. et al. 1998, \aj, 115, 2285

\bibitem[{{Manchester}(2006)}]{man06}
{Manchester}, R.~N. 2006, Chin. J. Astron. Astrophys. Suppl., 6, 139

\bibitem[{Mandelbaum {et~al.}(2005)Mandelbaum et al.}]{mhs+05}
{Mandelbaum}, R. et al. 2005, \mnras, 361, 1287

\bibitem[{{Marconi} \& {Hunt}(2003)}]{mh03}
{Marconi}, A. \& {Hunt}, L.~K. 2003, \apjl, 589, L21

\bibitem[{Masjedi {et~al.}(2006)Masjedi et al.}]{mhc+06}
{Masjedi}, M. et al. 2006, \apj, 644, 54

\bibitem[{{McIntosh} {et~al.}(2008){McIntosh}, {Guo}, {Hertzberg}, {Katz},
  {Mo}, {van den Bosch}, \& {Yang}}]{mgh+08}
{McIntosh}, D.~H., {Guo}, Y., {Hertzberg}, J., {Katz}, N., {Mo}, H.~J., {van
  den Bosch}, F.~C., \& {Yang}, X. 2008, \mnras, 388, 1537

\bibitem[{{McLure} \& {Dunlop}(2002)}]{md02}
{McLure}, R.~J. \& {Dunlop}, J.~S. 2002, \mnras, 331, 795

\bibitem[{{Merritt} \& {Ferrarese}(2001)}]{mf01}
{Merritt}, D. \& {Ferrarese}, L. 2001, \apj, 547, 140

\bibitem[{{Milosavljevi{\'c}} \& {Merritt}(2001)}]{mm01}
{Milosavljevi{\'c}}, M. \& {Merritt}, D. 2001, \apj, 563, 34

%\bibitem[{{Miyoshi} {et~al.}(1995){Miyoshi}, {Moran}, {Herrnstein},
%  {Greenhill}, {Nakai}, {Diamond}, \& {Inoue}}]{mmh+95}
%{Miyoshi}, M., {Moran}, J., {Herrnstein}, J., {Greenhill}, L., {Nakai}, N.,
%  {Diamond}, P., \& {Inoue}, M. 1995, \nat, 373, 127

\bibitem[{{Nelemans} {et~al.}(2001){Nelemans}, {Yungelson}, \& {Portegies
  Zwart}}]{nyp01}
{Nelemans}, G., {Yungelson}, L.~R., \& {Portegies Zwart}, S.~F. 2001, \aap,
  375, 890

\bibitem[{{Neuschaefer} {et~al.}(1997){Neuschaefer}, {Im}, {Ratnatunga},
  {Griffiths}, \& {Casertano}}]{nir+97}
{Neuschaefer}, L.~W., {Im}, M., {Ratnatunga}, K.~U., {Griffiths}, R.~E., \&
  {Casertano}, S. 1997, \apj, 480, 59

\bibitem[{{Owen} {et~al.}(1985){Owen}, {O'Dea}, {Inoue}, \& {Eilek}}]{ooi+85}
{Owen}, F.~N., {O'Dea}, C.~P., {Inoue}, M., \& {Eilek}, J.~A. 1985, \apjl, 294,
  L85

\bibitem[{{Oyaizu} {et~al.}(2008){Oyaizu}, {Lima}, {Cunha}, {Lin}, {Frieman},
  \& {Sheldon}}]{olc+08}
{Oyaizu}, H., {Lima}, M., {Cunha}, C.~E., {Lin}, H., {Frieman}, J., \&
  {Sheldon}, E.~S. 2008, \apj, 674, 768

\bibitem[{{Patton} \& {Atfield}(2008)}]{pa08}
{Patton}, D.~R. \& {Atfield}, J.~E. 2008, \apj, 685, 235

\bibitem[{{Patton} {et~al.}(2000){Patton}, {Carlberg}, {Marzke}, {Pritchet},
  {da Costa}, \& {Pellegrini}}]{pcm+00}
{Patton}, D.~R., {Carlberg}, R.~G., {Marzke}, R.~O., {Pritchet}, C.~J., {da
  Costa}, L.~N., \& {Pellegrini}, P.~S. 2000, \apj, 536, 153

\bibitem[{{Patton} {et~al.}(1997){Patton}, {Pritchet}, {Yee}, {Ellingson}, \&
  {Carlberg}}]{ppy+97}
{Patton}, D.~R., {Pritchet}, C.~J., {Yee}, H.~K.~C., {Ellingson}, E., \&
  {Carlberg}, R.~G. 1997, \apj, 475, 29

\bibitem[{Patton {et~al.}(2002)Patton et al.}]{ppc+02}
{Patton}, D.~R., et al. 2002, \apj, 565, 208

\bibitem[{{Peng} {et~al.}(2002){Peng}, {Ho}, {Impey}, \& {Rix}}]{phir02}
{Peng}, C.~Y., {Ho}, L.~C., {Impey}, C.~D., \& {Rix}, H.-W. 2002, \aj, 124, 266

\bibitem[{{Peters} \& {Mathews}(1963)}]{pm63}
{Peters}, P.~C. \& {Mathews}, J. 1963, Phys. Rev., 131, 435

\bibitem[{{Phinney}(2001)}]{phi01}
{Phinney}, E.~S. 2001, arXiv:astro-ph/0108028

\bibitem[{{Quinlan}(1996)}]{qui96}
{Quinlan}, G.~D. 1996, New Astron., 1, 35

\bibitem[{{Rajagopal} \& {Romani}(1995)}]{rr95}
{Rajagopal}, M. \& {Romani}, R.~W. 1995, \apj, 446, 543

\bibitem[{{Rines} {et~al.}(2007){Rines}, {Finn}, \& {Vikhlinin}}]{rfv07}
{Rines}, K., {Finn}, R., \& {Vikhlinin}, A. 2007, \apjl, 665, L9

\bibitem[{{Rodriguez} {et~al.}(2006){Rodriguez}, {Taylor}, {Zavala}, {Peck},
  {Pollack}, \& {Romani}}]{rtz+06}
{Rodriguez}, C., {Taylor}, G.~B., {Zavala}, R.~T., {Peck}, A.~B., {Pollack},
  L.~K., \& {Romani}, R.~W. 2006, \apj, 646, 49

\bibitem[{{Sazhin}(1978)}]{saz78}
{Sazhin}, M.~V. 1978, SvA, 22, 36

\bibitem[{{S{\'e}rsic}(1968)}]{ser68}
{S{\'e}rsic}, J.~L. 1968, {Atlas de Galaxias Australes} (Cordoba, Argentina:
  Observatorio Astronomico)

\bibitem[{{Sesana} {et~al.}(2006){Sesana}, {Haardt}, \& {Madau}}]{shm06}
{Sesana}, A., {Haardt}, F., \& {Madau}, P. 2006, \apj, 651, 392

\bibitem[{{Sesana} {et~al.}(2004){Sesana}, {Haardt}, {Madau}, \&
  {Volonteri}}]{shmv04}
{Sesana}, A., {Haardt}, F., {Madau}, P., \& {Volonteri}, M. 2004, \apj, 611,
  623

\bibitem[{{Sesana} {et~al.}(2008{\natexlab{a}}){Sesana}, {Vecchio}, \& {Colacino}}]{svc08}
{Sesana}, A., {Vecchio}, A., \& {Colacino}, C.~N. 2008{\natexlab{a}}, \mnras, 390, 192

\bibitem[{{Sesana} {et~al.}(2008{\natexlab{b}}){Sesana}, {Vecchio}, \& {Volonteri}}]{svv08} 
{Sesana}, A., {Vecchio}, A., \& {Volonteri}, M. 2008{\natexlab{b}}, arXiv:astro-ph/0809.3412

\bibitem[{{Sillanpaa} {et~al.}(1988){Sillanpaa}, {Haarala}, {Valtonen},
  {Sundelius}, \& {Byrd}}]{shv+88}
{Sillanpaa}, A., {Haarala}, S., {Valtonen}, M.~J., {Sundelius}, B., \& {Byrd},
  G.~G. 1988, \apj, 325, 628

\bibitem[{{Small} \& {Blandford}(1992)}]{sb92}
{Small}, T.~A. \& {Blandford}, R.~D. 1992, \mnras, 259, 725

\bibitem[{Stoughton {et~al.}(2002)Stoughton et al.}]{slb+02}
{Stoughton}, C., et al. 2002, \aj, 123, 485

\bibitem[{Strateva {et~al.}(2001)Strateva et al.}]{sik+01}
{Strateva}, I., et al. 2001, \aj, 122, 1861

\bibitem[{Strauss {et~al.}(2002)Strauss et al.}]{swl+02}
{ Strauss}, M.~A., et al. 2002, \aj, 124, 1810

\bibitem[{{Sudou} {et~al.}(2003){Sudou}, {Iguchi}, {Murata}, \&
  {Taniguchi}}]{simt03}
{Sudou}, H., {Iguchi}, S., {Murata}, Y., \& {Taniguchi}, Y. 2003, Science, 300,
  1263

\bibitem[{{Thorne}(1987)}]{tho87}
{Thorne}, K.~S. 1987, in Three Hundred Years of
  Gravitation, ed. S.~W. {Hawking} \& W. {Israel} 
(Cambridge: Cambridge University Press)

%\bibitem[{{Thorsett} \& {Dewey}(1996)}]{td96}
%{Thorsett}, S.~E. \& {Dewey}, R.~J. 1996, \prd, 53, 3468

\bibitem[{{Tundo} {et~al.}(2007){Tundo}, {Bernardi}, {Hyde}, {Sheth}, \&
  {Pizzella}}]{tbh+07}
{Tundo}, E., {Bernardi}, M., {Hyde}, J.~B., {Sheth}, R.~K., \& {Pizzella}, A.
  2007, \apj, 663, 53

\bibitem[{{Valtonen} {et~al.}(2008){Valtonen}, {Kidger}, {Lehto}, \&
  {Poyner}}]{vkl+08}
{Valtonen}, M., {Kidger}, M., {Lehto}, H., \& {Poyner}, G. 2008, \aap, 477, 407

\bibitem[{{van Dokkum}(2005)}]{van05}
{van Dokkum}, P.~G. 2005, \aj, 130, 2647

\bibitem[{{Woods} {et~al.}(1995){Woods}, {Fahlman}, \& {Richer}}]{wfr95}
{Woods}, D., {Fahlman}, G.~G., \& {Richer}, H.~B. 1995, \apj, 454, 32

\bibitem[{{Wu} \& {Keel}(1998)}]{wk98}
{Wu}, W. \& {Keel}, W.~C. 1998, \aj, 116, 1513

\bibitem[{{Wyithe} \& {Loeb}(2003)}]{wl03}
{Wyithe}, J.~S.~B. \& {Loeb}, A. 2003, \apj, 590, 691

\bibitem[{{Yee} \& {Ellingson}(1995)}]{ye95}
{Yee}, H.~K.~C. \& {Ellingson}, E. 1995, \apj, 445, 37

\bibitem[{{Yu}(2002)}]{yu02}
{Yu}, Q. 2002, \mnras, 331, 935

\bibitem[{{Zepf} \& {Koo}(1989)}]{zk89}
{Zepf}, S.~E. \& {Koo}, D.~C. 1989, \apj, 337, 34

\end{thebibliography}

\begin{center}
\begin{longtable}{rrrrrcccccccccl}

\caption{A list of luminous early-type galaxy pairs from the SDSS DR6}
%\label{tab1}\\
\\
\hline \hline 
\\ 
\multicolumn{1}{c}{No.} &\multicolumn{1}{c}{${\rm RA}_1$}&\multicolumn{1}{c}{${\rm Dec}_1$}&\multicolumn{1}{c}{${\rm RA}_2$}&\multicolumn{1}{c}{${\rm Dec}_2$}&$r_1$&$r_2$&$M_{r_1}$&$M_{r_2}$&D(kpc)&$a$&$r_{\rm res}$&$z$&$z$-flag &Comments\\
\multicolumn{1}{c}{(1)} &\multicolumn{1}{c}{(2)} &\multicolumn{1}{c}{(3)} &\multicolumn{1}{c}{(4)} &\multicolumn{1}{c}{(5)} & (6) & (7) & (8) &(9) &(10) &(11) &(12) &(13) &(14) &\multicolumn{1}{c}{(15)}\\
\hline\\
\endfirsthead

\multicolumn{15}{c}%
{\tablename\ \thetable{} -- continued} \vspace{1.mm}\\
\hline \hline 
\\
\multicolumn{1}{c}{No.} &\multicolumn{1}{c}{${\rm RA}_1$}&\multicolumn{1}{c}{${\rm Dec}_1$}&\multicolumn{1}{c}{${\rm RA}_2$}&\multicolumn{1}{c}{${\rm Dec}_2$}&$r_1$&$r_2$&$M_{r_1}$&$M_{r_2}$&D(kpc)&$a$&$r_{\rm res,r}$&$z$&$z$-flag&Comments\\
\multicolumn{1}{c}{(1)} &\multicolumn{1}{c}{(2)} &\multicolumn{1}{c}{(3)} &\multicolumn{1}{c}{(4)} &\multicolumn{1}{c}{(5)} & (6) & (7) & (8) &(9) &(10) &(11) &(12) &(13) &(14) &\multicolumn{1}{c}{(15)}\\
\hline\\
\endhead
\hline
\endfoot

\hline \hline
\endlastfoot

   1&   0.21881&  16.09328&   0.22045&  16.08902& 16.41& 16.88& -22.58& -21.51& 33.06&  0.25& 20.28& 0.1145& sp &non-merging \\
   2&   0.98566& -10.35133&   0.99133& -10.34565& 15.45& 15.77& -22.36& -23.42& 42.75&  1.47& 19.68& 0.0813& sp &merging     \\
   3&   1.40038&  15.93645&   1.40021&  15.94040& 16.17& 16.94& -22.53& -21.54& 28.83&  1.10& 20.01& 0.1148& sp &non-merging \\
   4&   1.40742&  16.22732&   1.40505&  16.22976& 16.86& 17.01& -21.64& -21.23& 23.36&  0.26& 20.56& 0.1095& sp &non-merging \\
   5&   1.91387&   0.69263&   1.91320&   0.68826& 15.95& 17.00& -23.01& -21.88& 32.05&  1.64& 18.83& 0.1141& sp &merging     \\
   6&   2.23632&  14.29741&   2.23641&  14.29035& 15.84& 15.94& -21.86& -22.36& 35.97&  0.00& 20.47& 0.0767& sp &non-merging \\
   7&   3.35190&   0.91775&   3.35490&   0.91536& 16.56& 16.98& -22.57& -21.98& 25.94&  0.00& 20.48& 0.1053& sp &non-merging \\
   8&   3.37760&   0.67847&   3.37124&   0.67352& 15.68& 16.30& -22.11& -21.73& 44.69&  0.17& 19.06& 0.0843& sp &non-merging \\
\end{longtable}
\tablecomments{
Col.(1): Number of pair; Col.(2): R.A.(J2000) of the first galaxy; Col.(3): Decl.(J2000) of the first galaxy; 
Col.(4): R.A.(J2000) of the second galaxy; Col.(5): Decl.(J2000) of the second galaxy; Col.(6): $r$-band magnitude of the first galaxy from SDSS pipeline; Col.(7): $r$-band magnitude of the second galaxy from SDSS pipeline; Col.(8): Our fitted $r$-band absolute magnitude of the first galaxy; Col.(9): Our fitted $r$-band absolute magnitude of the second galaxy; 
Col.(10): Separation of pair, in kpc; Col.(11): Asymmetry factor of pair; Col.(12): Residual magnitude of pair in $r$-band; 
Col.(13): Redshift of pair; Col.(14): ``sp'' for pair with spectroscopic
redshift; ``ph'' for pair with photometric redshift only; Col.(15): Merging
or non-merging classification.}

\end{center}

\end{document}